%% file: main.tex
\documentclass[runningheads]{llncs}
\usepackage[T1]{fontenc}
\input{preamble}

\begin{document}
\title{\hspace*{-0.71cm}\texttt{PEPSI}: Pathology-Enhanced Pulse-Sequence-Invariant\\Representations for Brain MRI}

\titlerunning{\texttt{PEPSI}}
\author{Peirong Liu \quad Oula Puonti \quad Annabel Sorby-Adams\\William T. Kimberly \quad Juan E. Iglesias}
\authorrunning{Liu et al.} 
\institute{Harvard Medical School and Massachusetts General Hospital}

\maketitle


\input{sec/abstract}

\input{sec/intro}

\input{sec/method/main}
\input{sec/exp/main}
\input{sec/con}

\clearpage
{\small
\bibliographystyle{splncs04}
\bibliography{reference}
}

\end{document}

%% file: preamble.tex
\usepackage{booktabs}

\usepackage{graphicx}
\usepackage{amsmath}
\usepackage{amsthm}
\usepackage{amssymb}  
\usepackage{pifont}
\newcommand{\cmark}{\ding{51}}%
\newcommand{\xmark}{\ding{55}}%

\usepackage{xcolor}
\usepackage{tikz}
\usepackage{wrapfig}
\usepackage{changepage}
\usepackage{tikz-3dplot}
\usepackage{tikzscale}
\usetikzlibrary{plotmarks}
\usepackage{pgfplots}
\usetikzlibrary{fadings}
\usetikzlibrary{tikzmark,calc} 
\usetikzlibrary{shapes.arrows}
\usetikzlibrary{decorations.pathreplacing, decorations.markings}
\usepgfplotslibrary{colorbrewer}
\usepgfmodule{nonlineartransformations} 
\makeatletter
\def\latticetilt{%
\pgf@xa=\pgf@x%
\pgf@ya=\pgf@y%
\pgfmathsetmacro{\myx}{\pgf@xa+\pgfkeysvalueof{/tikz/lattice/amplitude}*sin((\pgf@ya/\pgfkeysvalueof{/tikz/lattice/spacing})*360/\pgfkeysvalueof{/tikz/lattice/superlattice period})}%
\pgf@x=\myx pt%
\pgfmathsetmacro{\myy}{\pgf@ya+\pgfkeysvalueof{/tikz/lattice/amplitude}*sin((\pgf@xa/\pgfkeysvalueof{/tikz/lattice/spacing})*360/\pgfkeysvalueof{/tikz/lattice/superlattice period})}%
\pgf@y=\myy pt}

\usepackage[capbesideposition=right]{floatrow}

\xdefinecolor{navyblue}{RGB}{19 41 75}
\xdefinecolor{carolinablue}{RGB}{123 175 212}
\xdefinecolor{cb}{RGB}{123 175 212}
\xdefinecolor{flatgrey}{RGB}{162 170 173}
\xdefinecolor{matcha}{RGB}{183 224 176}

\definecolor{mygreen}{HTML}{DCEDC8}
\definecolor{lightgrey}{HTML}{E1E1E1}
\definecolor{darkblue}{HTML}{00275D}
\definecolor{mint}{HTML}{99EDC3}

\xdefinecolor{myyellow}{HTML}{FFE082}
\xdefinecolor{myblue}{RGB}{181, 222, 255}
\xdefinecolor{mypurple}{RGB}{202, 184, 255}
\xdefinecolor{myorange}{RGB}{255, 171, 145}
\xdefinecolor{tomato}{RGB}{255, 99, 71}

\definecolor{applegreen}{rgb}{0.4, 0.81, 0.0}
\definecolor{bittersweet}{rgb}{1.0, 0.2, 0.3}
\definecolor{mayablue}{rgb}{0.4, 0.5, 0.98}
\definecolor{citrine}{rgb}{1, 0.62, 0.14}
\definecolor{bubblegum}{rgb}{0.99, 0.76, 0.8}
\definecolor{olive}{rgb}{0.42, 0.56, 0.14}

\newcommand{\dashedLayer}[6]{
			\def\a{#1} 
			\def\b{0.02}
			\def\c{#2} 
			\def\t{#3} 
			\def\d{#4} 

			\draw[line width=0.15mm, dash pattern=on 2.25pt off 1.8pt](\c+\t,0,\d) -- (\c+\t,\a,\d) -- (\t,\a,\d);                                                      
			\draw[line width=0.15mm, dash pattern=on 2.25pt off 1.8pt](\t,0,\a+\d) -- (\c+\t,0,\a+\d) node[midway,below] {#6} -- (\c+\t,\a,\a+\d) -- (\t,\a,\a+\d) -- (\t,0,\a+\d); 
			\draw[line width=0.15mm, dash pattern=on 2.25pt off 1.8pt](\c+\t,0,\d) -- (\c+\t,0,\a+\d);
			\draw[line width=0.15mm, dash pattern=on 2.25pt off 1.8pt](\c+\t,\a,\d) -- (\c+\t,\a,\a+\d);
			\draw[line width=0.15mm, dash pattern=on 2.25pt off 1.8pt](\t,\a,\d) -- (\t,\a,\a+\d);
			
			\draw[line width=0.15mm] (\c+\t,0,\d) -- (\c+\t,\a,\d);
			\draw[line width=0.15mm] (\c+\t,0,\d) -- (\c+\t,0,\a+\d);
			\draw[line width=0.15mm] (\c+\t,\a,\d) -- (\c+\t,\a,\a+\d);
			\draw[line width=0.15mm] (\c+\t,0,\a+\d) -- (\c+\t,\a,\a+\d);
			
			\filldraw[#5] (\t+\b,\b,\a+\d) -- (\c+\t-\b,\b,\a+\d) -- (\c+\t-\b,\a-\b,\a+\d) -- (\t+\b,\a-\b,\a+\d) -- (\t+\b,\b,\a+\d); 
			\filldraw[#5] (\t+\b,\a,\a-\b+\d) -- (\c+\t-\b,\a,\a-\b+\d) -- (\c+\t-\b,\a,\b+\d) -- (\t+\b,\a,\b+\d);
			\ifthenelse {\equal{#5} {}}
			{} 
			{\filldraw[#5] (\c+\t,\b,\a-\b+\d) -- (\c+\t,\b,\b+\d) -- (\c+\t,\a-\b,\b+\d) -- (\c+\t,\a-\b,\a-\b+\d);} 
		}

\newcommand{\networkLayer}[6]{
			\def\a{#1} 
			\def\b{0.02}
			\def\c{#2} 
			\def\t{#3} 
			\def\d{#4} 

			\draw[line width=0.15mm](\c+\t,0,\d) -- (\c+\t,\a,\d) -- (\t,\a,\d);                                                      
			\draw[line width=0.15mm](\t,0,\a+\d) -- (\c+\t,0,\a+\d) node[midway,below] {#6} -- (\c+\t,\a,\a+\d) -- (\t,\a,\a+\d) -- (\t,0,\a+\d); 
			\draw[line width=0.15mm](\c+\t,0,\d) -- (\c+\t,0,\a+\d);
			\draw[line width=0.15mm](\c+\t,\a,\d) -- (\c+\t,\a,\a+\d);
			\draw[line width=0.15mm](\t,\a,\d) -- (\t,\a,\a+\d);

			\filldraw[#5] (\t+\b,\b,\a+\d) -- (\c+\t-\b,\b,\a+\d) -- (\c+\t-\b,\a-\b,\a+\d) -- (\t+\b,\a-\b,\a+\d) -- (\t+\b,\b,\a+\d); 
			\filldraw[#5] (\t+\b,\a,\a-\b+\d) -- (\c+\t-\b,\a,\a-\b+\d) -- (\c+\t-\b,\a,\b+\d) -- (\t+\b,\a,\b+\d);
			\ifthenelse {\equal{#5} {}}
			{} 
			{\filldraw[#5] (\c+\t,\b,\a-\b+\d) -- (\c+\t,\b,\b+\d) -- (\c+\t,\a-\b,\b+\d) -- (\c+\t,\a-\b,\a-\b+\d);} 
		}
		
\usepackage[linesnumbered,ruled,vlined]{algorithm2e}

\SetCommentSty{mycommfont}
\SetKwInput{Input}{Input}                
\SetKwInput{Output}{Output}         
\SetKwInput{Dataset}{Dataset}     
\SetKwInput{Parameters}{Parameters}  
\SetKwInput{Settings}{Settings}      
\SetKwInput{Initialization}{Initialization}      
\SetAlgorithmName{Alg.}{Alg.}{List of Algorithms} 
\makeatletter
\newcommand{\algorithmfootnote}[2][\footnotesize]{%
  \let\old@algocf@finish\@algocf@finish
  \def\@algocf@finish{\old@algocf@finish
    \leavevmode\rlap{\begin{minipage}{\linewidth}
    #1#2
    \end{minipage}}%
  }%
}
\makeatother

\usepackage{floatrow} 
\usepackage{makecell}

\usepackage{booktabs}
\usepackage{multirow}
\newcolumntype{P}[1]{>{\centering\arraybackslash}p{#1}}

\newtheoremstyle{bfnote}%
  {}{}
  {\itshape}{}
  {\bfseries}{.}
  { }{\thmname{#1}\thmnumber{ #2}\thmnote{ (#3)}}
\theoremstyle{bfnote}

\usepackage{url}

\pgfplotsset{compat=1.18}


\usepackage[pagebackref,breaklinks,colorlinks]{hyperref} 
  
\usepackage[capitalize]{cleveref}
\crefname{section}{Sec.}{Secs.}
\Crefname{section}{Section}{Sections}
\Crefname{table}{Table}{Tables}
\crefname{table}{Tab.}{Tabs.}

%% file: sec/abstract.tex
\begin{abstract} 
Remarkable progress has been made by data-driven machine-learning methods in the analysis of MRI scans. However, most existing MRI analysis approaches are crafted for specific MR pulse sequences (MR contrasts) and usually require nearly isotropic acquisitions. This limits their applicability to diverse real-world clinical data, where scans commonly exhibit variations in appearances due to being obtained with varying sequence parameters, resolutions, and orientations -- especially in the presence of pathology. In this paper, we propose \texttt{PEPSI}, the first pathology-enhanced, and pulse-sequence-invariant feature representation learning model for brain MRI. \texttt{PEPSI} is trained entirely on synthetic images with a novel pathology encoding strategy, and enables co-training across datasets with diverse pathologies and missing modalities. Despite variations in pathology appearances across different MR pulse sequences or the quality of acquired images (e.g., resolution, orientation, artifacts, etc), \texttt{PEPSI} produces a high-resolution image of reference contrast (MP-RAGE) that captures anatomy, along with an image specifically highlighting the pathology. Our experiments demonstrate \texttt{PEPSI}'s remarkable capability for image synthesis compared with the state-of-the-art, contrast-agnostic synthesis models, as it accurately reconstructs anatomical structures while differentiating between pathology and normal tissue. We further illustrate the efficiency and effectiveness of \texttt{PEPSI} features for downstream pathology segmentations on five public datasets covering white matter hyperintensities and stroke lesions. Code is available at \href{https://github.com/peirong26/PEPSI}{https://github.com/peirong26/PEPSI}.
\end{abstract}

%% file: sec/intro.tex
\section{Introduction}
\label{sec: intro}  
Recent learning based methods have enabled considerably more rapid and accurate image analysis of brain magnetic resonance imaging (MRI)~\cite{Iglesias2023SynthSRAP}, which provides precise and adjustable soft-tissue contrast via noninvasive, in vivo imaging of the human brain~\cite{BrantZawadzki1992MPRA}. Nevertheless, the majority of current MRI analysis approaches are tailored to particular MR pulse sequences (MR contrast), and often rely on nearly isotropic acquisitions. Consequently, sharp declines in performance frequently occur when voxel size and anisotropy increase, or when applied to a contrast different from the one used during training~\cite{wang2018deep}. This compromises model generalizability and leads to extra data collection and training efforts when dealing with new datasets. Leveraging synthetic data, recent contrast-agnostic models~\cite{Iglesias2020JointSA,Liu2021YETI,Liu2022SONATA,Iglesias2023SynthSRAP,Billot2021SynthSegSO,Hoffmann2020SynthMorphLC,Laso2023WMH} demonstrate remarkable performance and largely broaden the scope of model applicability to the diverse clinical acquisition protocols. However, these models are confined to the specific tasks they were trained for and cannot be readily adapted to other tasks.

Meanwhile, task-agnostic foundation models~\cite{Bommasani2021OnTO,Awais2023FoundationalMD} in general computer vision and natural language processing have experienced notable success, driven by the fast growth of large-scale datasets~\cite{Brown2020LanguageMA,Chowdhery2022PaLMSL,Kirillov2023SegmentA}. Nonetheless, the development of foundation models in  medical imaging have been hindered by the lack of large-scale datasets (in many domains), variations in acquisition protocols and processing pipelines, and privacy constraints.  MONAI~\cite{cardoso2022monai} provides pre-trained models for diverse tasks, but they generally are highly task-oriented and contrast-sensitive. Zhou et al.~\cite{Zhou2023AFM} proposed a medical foundation model, which is specifically designed for the detection of eye and systemic health conditions from retinal scans, yet this model is limited to the modalities of color fundus photography and optical coherence tomography. AI generalist systems~\cite{Moor2023FoundationMF,Singhal2022LargeLM,Tu2023TowardsGB} have shown superiority in biomedical tasks (e.g., visual question answering, image classification, radiology report generation and summarizing), but mostly within the vision-language context. CIFL~\cite{chua2023contrast} was designed for task-agnostic feature representations, yet it has only been demonstrated in 2D, and exclusively relies on contrastive learning, insufficient in surpassing task-specific models in downstream applications~\cite{Liu2023BrainID}. Recently, Liu et al.~\cite{Liu2023BrainID} proposed Brain-ID, which extracts contrast-agnostic features for brain MRI, and achieves state-of-the-art performance in various fundamental medical imaging tasks including reconstruction, segmentation, and super-resolution. 
However, Brain-ID exclusively focuses on healthy-appearing anatomy and lacks the capacity to model pathologies (\cref{fig: comp}).

\vspace{0.06cm}
In this paper, we introduce \texttt{PEPSI}, the first pulse-sequence-invariant feature representation learning approach specifically designed to emphasize pathology. \texttt{PEPSI} is trained on synthetic data encoded with pathology, and can be directly applied to real images featuring various types of pathology. 
\vspace{-0.07cm}
\begin{itemize}
    \item[1)] We introduce a data generator that synthesizes images incorporating augmented pathologies across \textit{any} combination of deformation, pulse sequence, resolution, orientation, artifacts, etc., thus circumventing the limitations of real data, which are often confined to the acquired pulse sequence~(\cref{fig: augment}). 
    \vspace{0.1cm}
    \item[2)] We design a feature learning framework guided by MP-RAGE and FLAIR scans, which balances anatomy and pathology. Furthermore, \texttt{PEPSI} bridges the gaps of pathologies across datasets via our proposed implicit pathology supervision, and enables co-training across datasets with different pathology types and potentially missing modalities~(\cref{sec: framework}). 
    \vspace{0.1cm}
    \item[3)] We conduct comprehensive evaluations on image synthesis and pathology segmentation. \texttt{PEPSI} exhibits: \textit{(i)}~a remarkable capability to synthesize images with missing modalities while simultaneously capturing various pathologies~(\cref{fig: comp}); \textit{(ii)}~superior efficiency and effectiveness on downstream pathology segmentation across five public datasets, covering modalities of T1w and FLAIR, with white matter hyperintensity (WMH) and stroke lesions~(\cref{tab: seg}).
\end{itemize}

%% file: sec/method/main.tex
\vspace{-0.2cm}
\section{Approach}
\label{sec: method}
\vspace{-0.22cm}

As mentioned earlier, sourcing large-scale datasets with high-quality and diverse contrasts for brain MRI remains challenging. Recent works~\cite{Billot2021SynthSegSO,Hoopes2022SynthStripSF,Iglesias2023SynthSRAP,Liu2023BrainID} have addressed this issue by utilizing anatomy labels to simulate data, thereby enriching the learning space. However, all their training data generators are \textit{solely} based on brain anatomy and lack prior information on any potential pathology. Instead, we seek to synthesize data that \textit{emphasizes} pathologies (\cref{sec: generator}), and encourage the model to \textit{distinguish} between normal and abnormal regions in the resulting features (\cref{sec: framework}), facilitating the transmission of valuable information for downstream pathology detection and segmentation tasks.


\input{sec/method/generator}
\input{sec/method/trainer}


%% file: sec/method/generator.tex
\vspace{-0.4cm}
\subsection{Generating Pathology-encoded Training Data}
\label{sec: generator}
\vspace{-0.2cm}

\input{sec/method/fw_gen}

\texttt{PEPSI} leverages neuroanatomical labels and pathology segmentations to generate contrast-diverse data while \textit{simultaneously} emphasizing pathology.

\vspace{-0.5cm}
\subsubsection{Anomaly Probabilities:}
\label{sec: abnomaly}
We construct a proxy for soft anomaly maps ($P$) from the intensities of an image ($I$) using \textit{a priori} knowledge of its expected appearance, conditioned on the MR contrast, and the nature of the expected lesions (e.g., white matter lesions, and stroke):
\vspace{-0.2cm}
\begin{equation} 
P(x) =
    \begin{cases}
        0\,, & x \notin \Omega_P \\ 
        1- (I(x) - I_{\min}) / (I_{\max} - I_{\min})\,, & x \in \Omega_P\,, ~I \in \{\text{T1w}\} \\
        (I(x) - I_{\min}) / (I_{\max} - I_{\min})\,, & x \in \Omega_P\,, ~I \in \{\text{T2w, FLAIR}\} 
    \end{cases}
    \label{eq: prob}
\vspace{-0.15cm}
\end{equation} 
where $\Omega_P$ refers to the pathological region, $I_{\max}$ ($I_{\min}$) is the regional maximum (minimum) image intensities: $I_{\max} = \max_{x\in\Omega_P}I(x),\, I_{\min} = \min_{x\in\Omega_P}I(x)$.


\vspace{-0.45cm}
\subsubsection{Pathology-encoded Contrast:}
\label{sec: contrast}

To generate images with complex brain structures, we leverage anatomy labels following~\cite{Liu2023BrainID}. As shown in \cref{fig: augment}, a random deformation field ($\phi$) is first generated, comprising linear and non-linear transformations~\cite{Iglesias2020JointSA,Liu2023BrainID}. After the anatomy labels ($L$) and anomaly probabilities ($P$) are deformed by $\phi$, we generate the pathology-encoded images via two steps: 

\vspace{0.05cm}
\noindent \textit{(i)} \textit{``Anomaly-free'' image ($S_0$)}: We begin with randomly sampling intensities on the brain anatomy labels, where the regional intensities are generated by independently sampling a Gaussian distribution for each labeled region~\cite{Liu2023BrainID}.  

\noindent \textit{(ii)} \textit{Pathology enhancement}: We incorporate the anomaly probabilities into the ``anomaly-free'' image ($S_0$) to produce a pathology-encoded image ($S$) -- again, using \textit{a priori} knowledge of the modality. This is conditioned on the direction of intensities from white to gray matter in $S_0$: $S(x) = S_0(x) + \Delta S(x) * p(x)\,$,
\vspace{-0.2cm}
\begin{equation} 
\hspace{-0.2cm}
\text{s.t.}~\Delta S(x) \sim    
    \begin{cases} 
        \{ 0 \} \,, & x \notin \Omega_{\phi \circ P}\\ 
        \mathcal{N}(-\mu_{\text{w}} / 2 ,\, \mu_{\text{w}} / 2 ) \,,& x \in \Omega_{\phi \circ P}\,, \mu_{\text{w}} > \mu_{\text{g}}\\
        \mathcal{N}(\mu_{\text{w}} / 2 ,\, \mu_{\text{w}} / 2 ) \,, & x \in \Omega_{\phi \circ P}\,, \mu_{\text{w}} \leq \mu_{\text{g}} \\ 
    \end{cases}
    \label{eq: contrast} 
\vspace{-0.24cm}
\end{equation}
$\mu_{\text{w}}$ ($\mu_{\text{g}}$) is the mean value of white (gray) matter intensities in $S_0$. A higher $\mu_{\text{w}}$ resembles T1w, where pathologies appear darker; A lower $\mu_{\text{w}}$ resembles T2w or FLAIR, where pathologies are typically brighter. (See the dashed box in \cref{fig: augment}.)

As shown in \cref{fig: augment}, the pathology-encoded images ($S$) further undergo the corruption pipeline~\cite{Iglesias2023SynthSRAP}, which introduces various levels of resolution, noises and scanning artifacts commonly encountered in clinical protocols. 

%% file: sec/method/fw_gen.tex
\begin{figure}[t]
\centering
\resizebox{\textwidth}{!}{
	
	\begin{tikzpicture}[lattice/.cd,spacing/.initial=4,superlattice
  period/.initial=12,amplitude/.initial=2]
	\pgfmathsetmacro{\cubex}{0.29*3}
	\pgfmathsetmacro{\cubey}{0.29*3}
	\pgfmathsetmacro{\cubez}{0.028*3}
	


	\pgfmathsetmacro{\shift}{-2.5}
	\foreach \i in {1.3}
	{
	\draw[black,fill=gray!30, line width = 0.02mm] (\i+\shift+0.4*3,0.5*3,0.5*3) -- ++(-\cubex,0,0) -- ++(0,-\cubey,0) -- ++(\cubex,0,0) -- cycle;
	\draw[black,fill=gray!35, line width = 0.02mm] (\i+\shift+0.4*3,0.5*3,0.5*3) -- ++(0,0,-\cubez) -- ++(0,-\cubey,0) -- ++(0,0,\cubez) -- cycle;
	\draw[black,fill=gray!35, line width = 0.02mm] (\i+\shift+0.4*3,0.5*3,0.5*3) -- ++(-\cubex,0,0) -- ++(0,0,-\cubez) -- ++(\cubex,0,0) -- cycle;
	\node at (\i+\shift+0.409*3, 0.509*3, 0.9*3) {\includegraphics[width=0.071\textwidth]{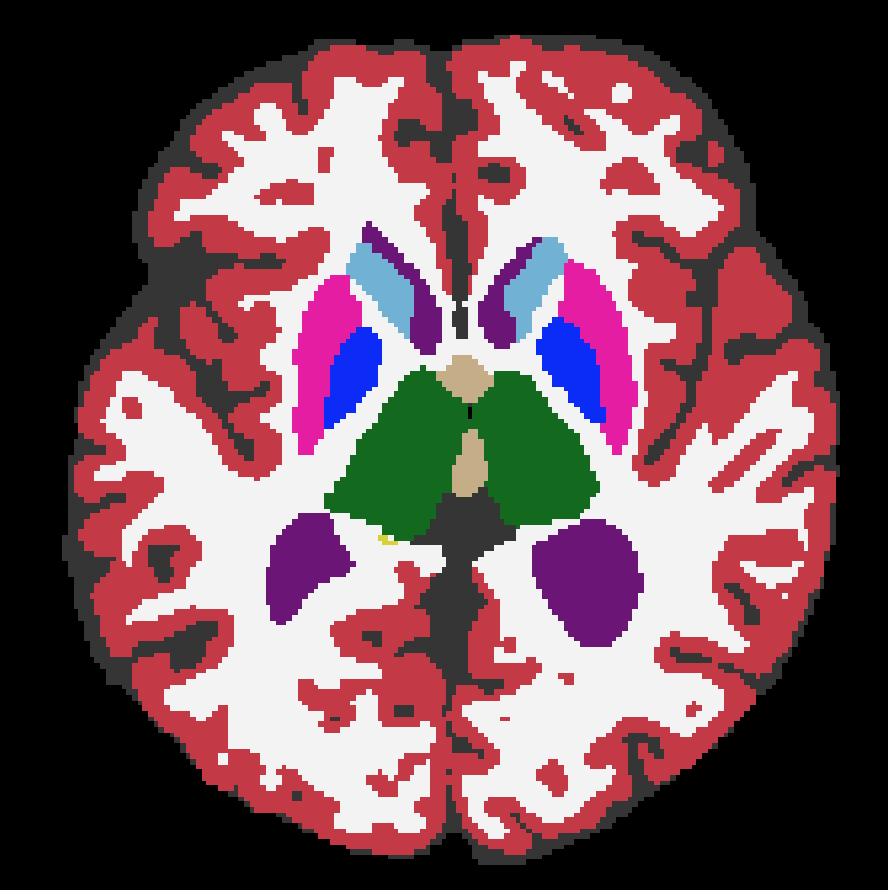}};
	\node at (\i+\shift+0.409*3, 0.95, 0.9*3) {\tiny $L$};
	}
 
	\foreach \i in {1.995}
	{
	\draw[black,fill=gray!30, line width = 0.02mm] (\i+\shift-0.07*3,-0.17*3,-0.12*3) -- ++(-\cubex,0,0) -- ++(0,-\cubey,0) -- ++(\cubex,0,0) -- cycle;
	\draw[black,fill=gray!35, line width = 0.02mm] (\i+\shift-0.07*3, -0.17*3,-0.12*3) -- ++(0,0,-\cubez) -- ++(0,-\cubey,0) -- ++(0,0,\cubez) -- cycle;
	\draw[black,fill=gray!35, line width = 0.02mm] (\i+\shift-0.07*3,-0.17*3,-0.12*3) -- ++(-\cubex,0,0) -- ++(0,0,-\cubez) -- ++(\cubex,0,0) -- cycle;	
	\node at (\i+\shift-0.75, -0.35*3, -0.212*3) {\includegraphics[width=0.071\textwidth]{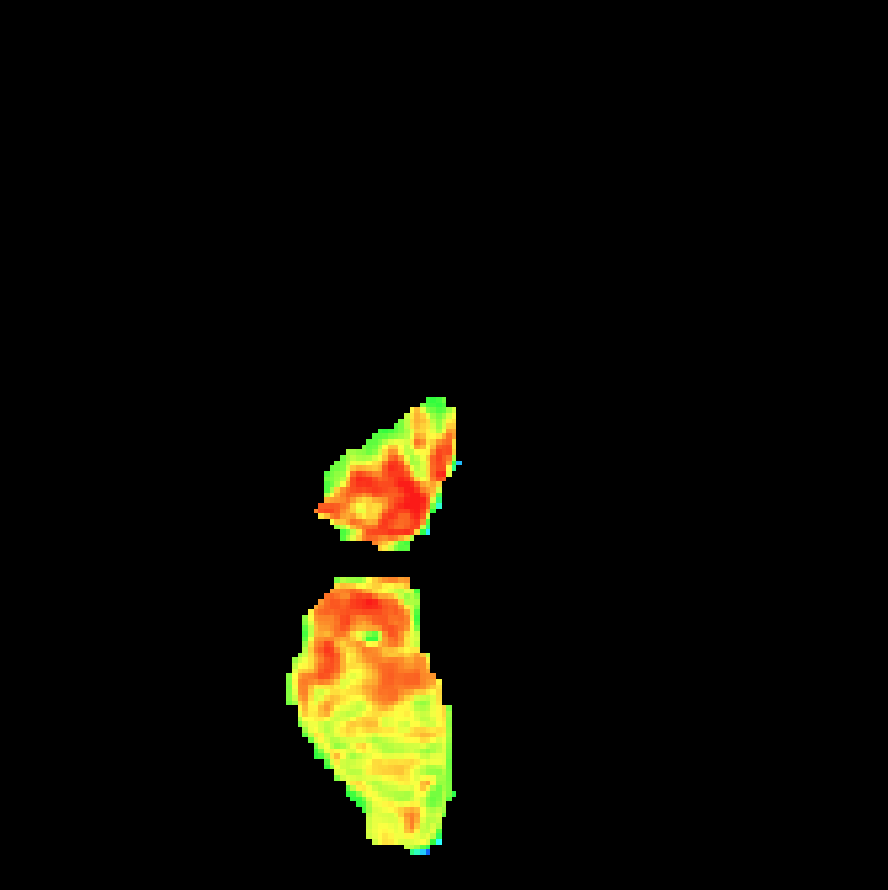}}; 
	\node at (\i+\shift-0.75, -1.64, -0.212*3) {\tiny $P$~(\cref{eq: prob})};
	}

	\pgfmathsetmacro{\shift}{0.5}
	\draw[-latex] (-1.15+0.6+\shift, 0.05, 0.25*3) -- (0.1+0.8+\shift, 0.05, 0.25*3); 
	\node at (-1.0+1.15+\shift, -0.11, 0.25*3) {\tiny \text{Deformation}};

	\node at (-1.0+1.15+\shift, -0.55, 0.25*3) {\includegraphics[width=0.08\textwidth]{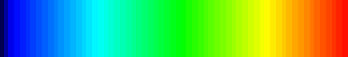}}; 
	\node at (-1.0+1.15-0.62+\shift, -0.55, 0.25*3) {\tiny \text{0}}; 
	\node at (-1.0+1.15+0.62+\shift, -0.55, 0.25*3) {\tiny \text{1}}; 
	\node at (-1.0+1.15+\shift, -0.8, 0.25*3) {\tiny \text{Anomaly Prob}};

	\pgfmathsetmacro{\shift}{0.3}
	\pgfmathsetmacro{\dx}{-5.5}
	\pgfmathsetmacro{\dy}{2.7}
        
        \begin{scope}[xshift=-5]
            \pgftransformnonlinear{\latticetilt}
            \draw[orange!80,step=0.08,thin] (0.15,0.15) grid (0.88,0.88);
            \draw[->, black] (0.15,0.15) -- +(0.85,0);
            \draw[->, black] (0.15,0.15) -- +(0,0.85);
        \end{scope} 
    
	\draw[->, line width = 0.2mm, color=orange!80] (5.6+\dx+\shift, -2.55+\dy) -- (5.6+\dx+\shift, -2.87+\dy);
 
	\node at (5.45+\dx+\shift, -2.7+\dy) {\tiny $\phi$}; 
	

	\foreach \i in {1.305}
	{
	\draw[black,fill=gray!30, line width = 0.02mm] (\i+\shift+0.4*3,0.5*3,0.5*3) -- ++(-\cubex,0,0) -- ++(0,-\cubey,0) -- ++(\cubex,0,0) -- cycle;
	\draw[black,fill=gray!35, line width = 0.02mm] (\i+\shift+0.4*3,0.5*3,0.5*3) -- ++(0,0,-\cubez) -- ++(0,-\cubey,0) -- ++(0,0,\cubez) -- cycle;
	\draw[black,fill=gray!35, line width = 0.02mm] (\i+\shift+0.4*3,0.5*3,0.5*3) -- ++(-\cubex,0,0) -- ++(0,0,-\cubez) -- ++(\cubex,0,0) -- cycle;
	\node at (\i+\shift+0.409*3, 0.509*3, 0.9*3) {\includegraphics[width=0.071\textwidth]{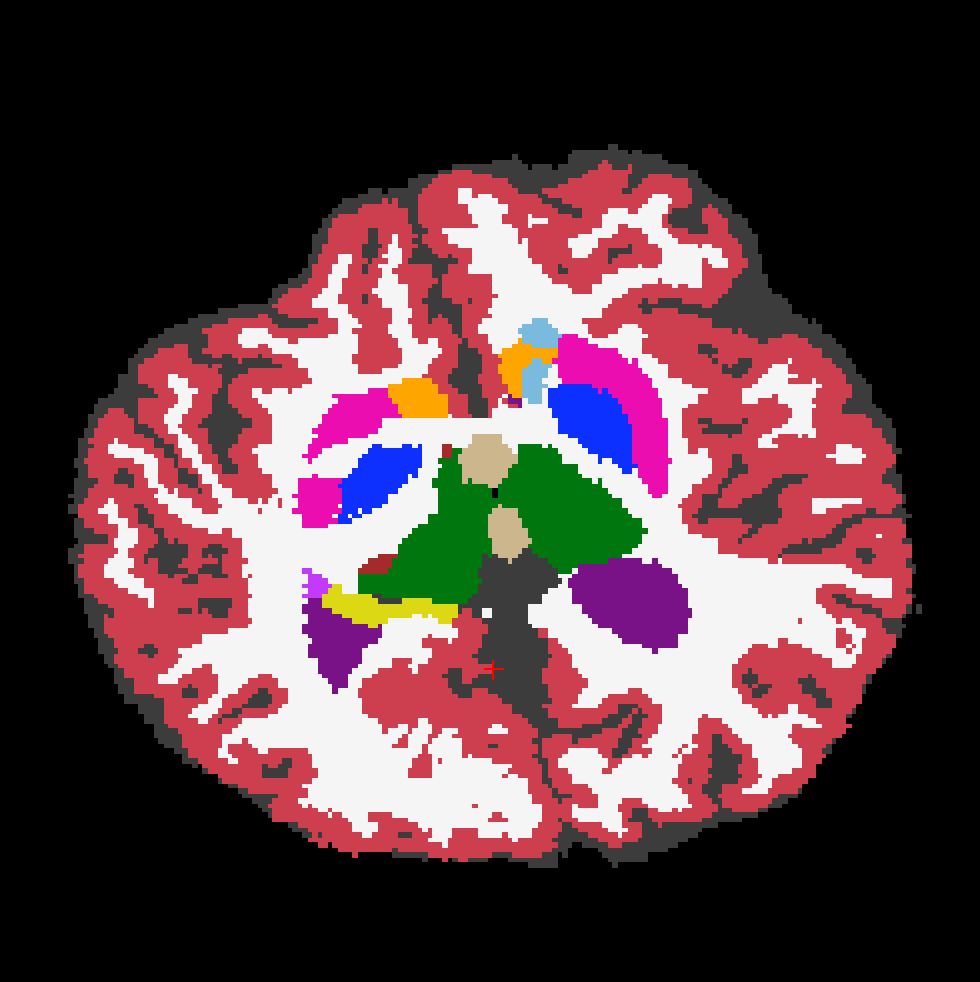}}; 
	\node at (\i+\shift+0.409*3, 0.95, 0.9*3) {\tiny $\phi \circ L$};
	}
 
	\foreach \i in {2.}
	{
	\draw[black,fill=gray!30, line width = 0.02mm] (\i+\shift-0.07*3,-0.17*3,-0.12*3) -- ++(-\cubex,0,0) -- ++(0,-\cubey,0) -- ++(\cubex,0,0) -- cycle;
	\draw[black,fill=gray!35, line width = 0.02mm] (\i+\shift-0.07*3,-0.17*3,-0.12*3) -- ++(0,0,-\cubez) -- ++(0,-\cubey,0) -- ++(0,0,\cubez) -- cycle;
	\draw[black,fill=gray!35, line width = 0.02mm] (\i+\shift-0.07*3,-0.17*3,-0.12*3) -- ++(-\cubex,0,0) -- ++(0,0,-\cubez) -- ++(\cubex,0,0) -- cycle;	
	\node at (\i+\shift-0.75, -0.35*3, -0.212*3) {\includegraphics[width=0.071\textwidth]{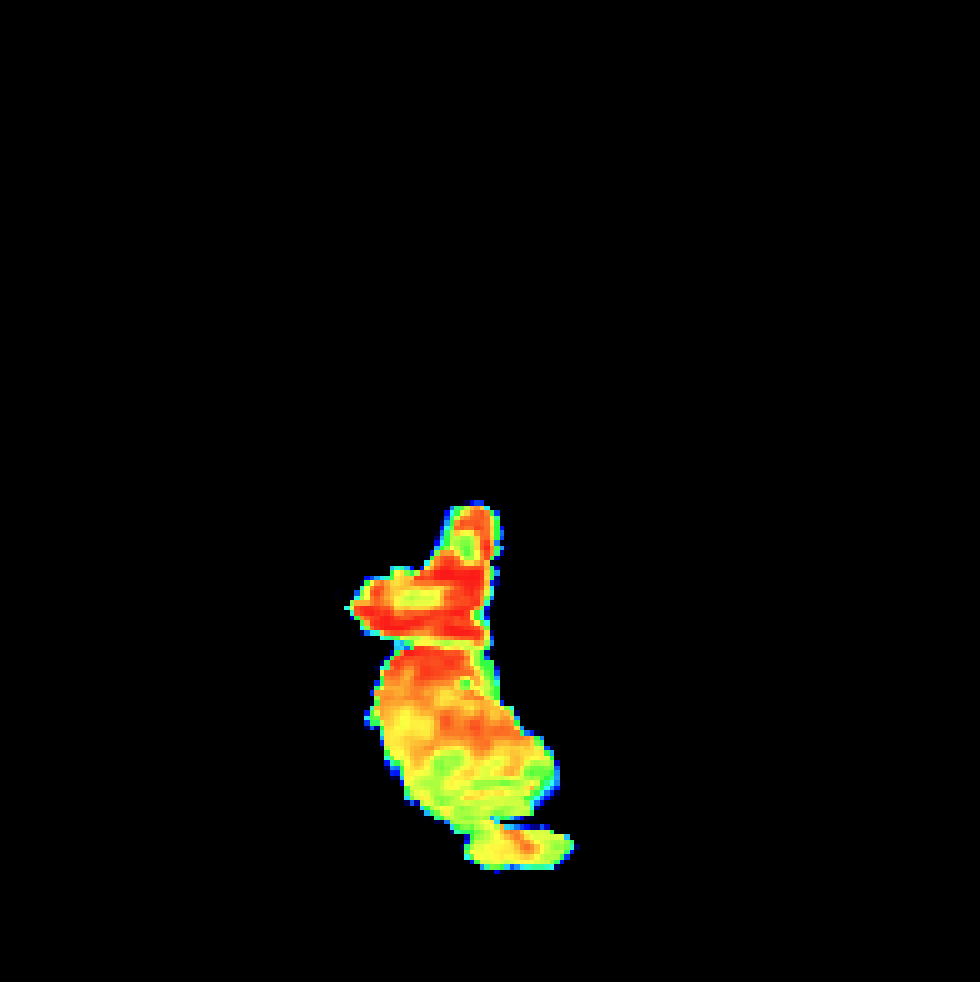}};
	\node at (\i+\shift-0.75, -1.64, -0.212*3) {\tiny $\phi \circ P$};
	}

	\node at (2.+1.04+\shift, 0.85, 0.25*3) {\includegraphics[width=0.1\textwidth]{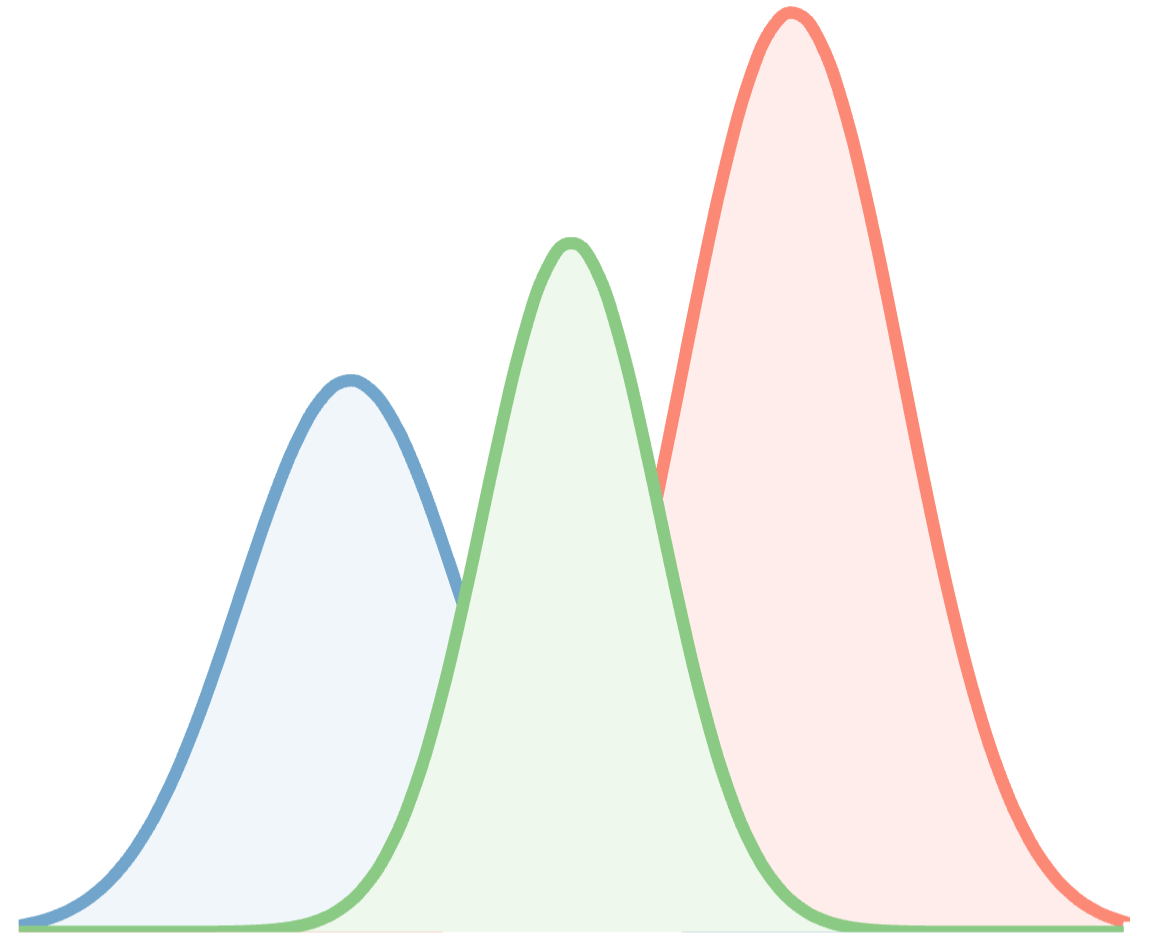}};
 
	\pgfmathsetmacro{\shift}{3.25}
 
	\node at (-1.0+1.15+\shift, -0.11, 0.25*3) {\tiny \text{Pathology-}};
	\node at (-1.0+1.09+\shift, -0.31, 0.25*3) {\tiny \text{enhanced}};
	\node at (-1.0+1.08+\shift, -0.51, 0.25*3) {\tiny \text{Sampling}};
	\node at (-1.0+1.05+\shift, -0.73, 0.25*3) {\tiny (\cref{eq: contrast})};

	\draw[-latex] (-1.15+0.6+\shift, 0.05, 0.25*3) -- (0.1+0.8+\shift, 0.05, 0.25*3);
	
	\draw[->, line width = 0.2mm, color=cb!90] (4.94+\dx+\shift, -2.63+\dy) -- (4.94+\dx+\shift, -2.87+\dy);
	\draw[->, line width = 0.2mm, color=matcha!150] (5.3+\dx+\shift, -2.63+\dy) -- (5.3+\dx+\shift, -2.87+\dy);
	\draw[->, line width = 0.2mm, color=tomato!70] (5.66+\dx+\shift, -2.63+\dy) -- (5.66+\dx+\shift, -2.87+\dy);

	
	\pgfmathsetmacro{\dx}{0.} 
	\pgfmathsetmacro{\ddx}{-0.}
	\pgfmathsetmacro{\ddy}{-0.4}
        \foreach \dy in {4.}
        {
	\draw[dashed, color = flatgrey!80, line width=0.4mm] (4+\dx, -5+\dy+\ddy) -- (8.8+\dx+\ddx, -5+\dy+\ddy) -- (8.8+\dx+\ddx, -3.+\dy) -- (4+\dx, -3.+\dy) -- (4+\dx, -5+\dy+\ddy); 
        }
        
	\node at (1.5+1.66+\shift, 0.8+0.22, 0.25*3) {\tiny \text{w/ random Corruption $+$ Resampling}};
        
	\pgfmathsetmacro{\cubeza}{0.05*3}
	\pgfmathsetmacro{\cubezb}{0.02*3}
	\pgfmathsetmacro{\shift}{4.2}
	\foreach \i in {1}
	{
	\draw[black,fill=gray!30, line width = 0.02mm] (\i+\shift-0.1*3,0.*3,-0.1*3) -- ++(-\cubex,0,0) -- ++(0,-\cubey,0) -- ++(\cubex,0,0) -- cycle;
	\draw[black,fill=gray!35, line width = 0.02mm] (\i+\shift-0.1*3, 0.*3,-0.1*3) -- ++(0,0,-\cubezb) -- ++(0,-\cubey,0) -- ++(0,0,\cubezb) -- cycle;
	\draw[black,fill=gray!35, line width = 0.02mm] (\i+\shift-0.1*3,0.*3,-0.1*3) -- ++(-\cubex,0,0) -- ++(0,0,-\cubezb) -- ++(\cubex,0,0) -- cycle;
	
	\node at (\i+\shift-0.28*3, -0.18*3, -0.19*3) {\includegraphics[width=0.071\textwidth]{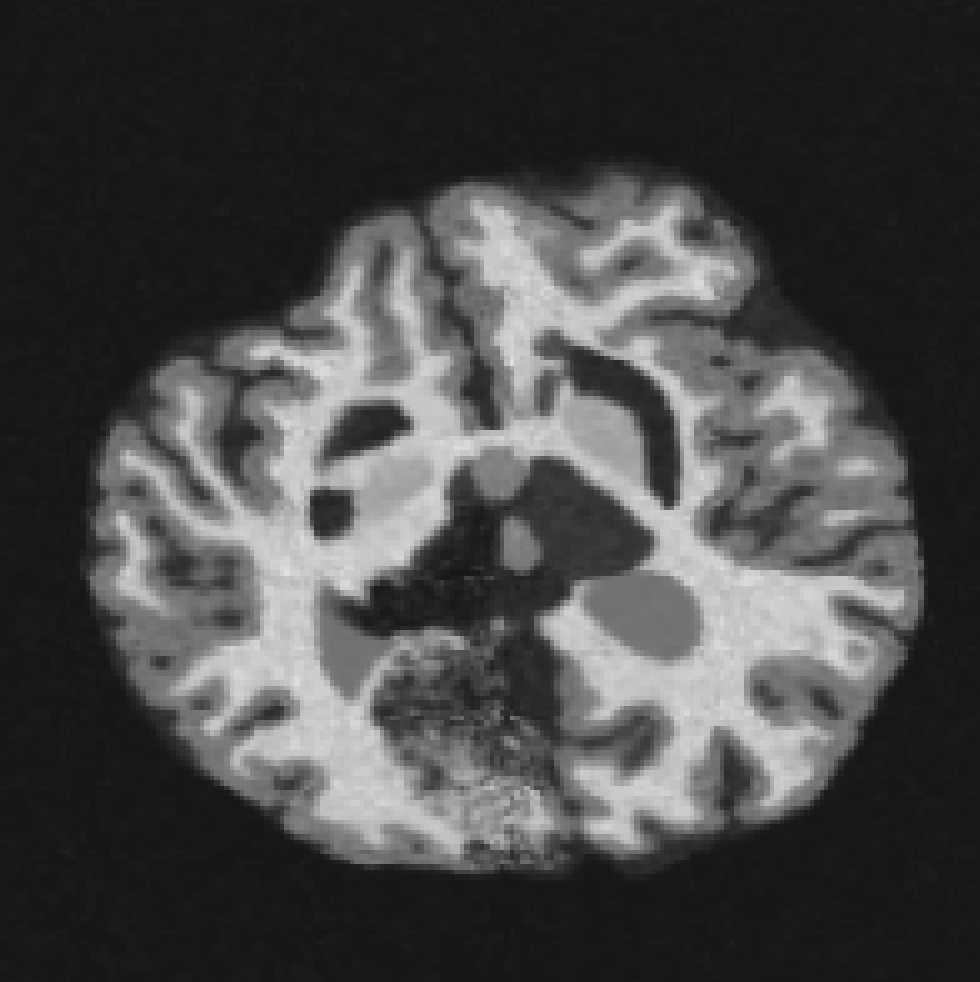}};
	\node at (\i+\shift+0.65, 0.23+1.3, 3.1) {\tiny $S_1$};
	}

	\pgfmathsetmacro{\cubeza}{0.03*3}
	\pgfmathsetmacro{\cubezb}{0.04*3}
	\pgfmathsetmacro{\shift}{4.2}
	\foreach \i in {2}
	{
	\draw[black,fill=gray!30, line width = 0.02mm] (\i+\shift-0.1*3,0.*3,-0.1*3) -- ++(-\cubex,0,0) -- ++(0,-\cubey,0) -- ++(\cubex,0,0) -- cycle;
	\draw[black,fill=gray!35, line width = 0.02mm] (\i+\shift-0.1*3, 0.*3,-0.1*3) -- ++(0,0,-\cubezb) -- ++(0,-\cubey,0) -- ++(0,0,\cubezb) -- cycle;
	\draw[black,fill=gray!35, line width = 0.02mm] (\i+\shift-0.1*3,0.*3,-0.1*3) -- ++(-\cubex,0,0) -- ++(0,0,-\cubezb) -- ++(\cubex,0,0) -- cycle;
	
	\node at (\i+\shift-0.28*3, -0.18*3, -0.19*3) {\includegraphics[width=0.071\textwidth]{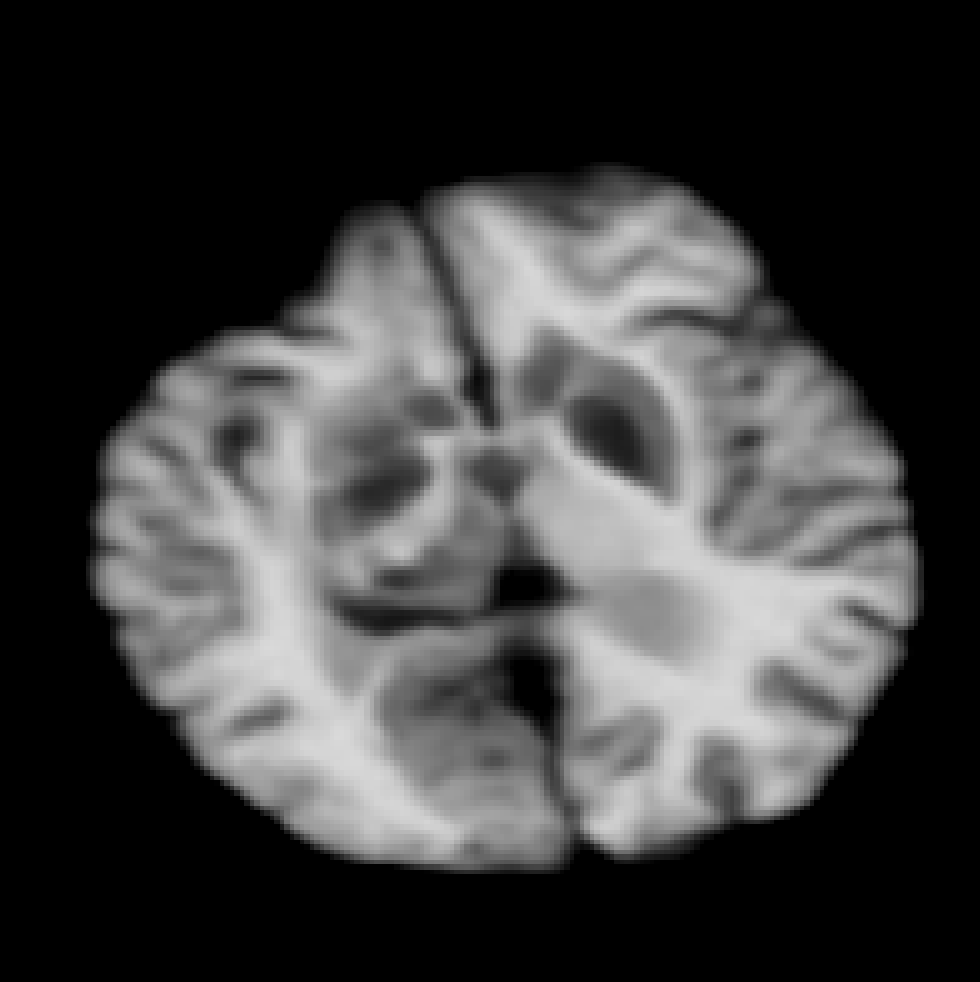}};
	\node at (\i+\shift+0.65, 0.23+1.3, 3.1) {\tiny $S_2$};
 
	\draw[black,fill=black] (\i+\shift+0.36, 0.03, 0.25*3) circle (0.4pt);
	\draw[black,fill=black] (\i+\shift+0.46, 0.03, 0.25*3) circle (0.4pt);
	\draw[black,fill=black] (\i+\shift+0.56, 0.03, 0.25*3) circle (0.4pt);
	}
 
    \draw [decorate,decoration={brace,amplitude=5pt,mirror,raise=6ex},line width=1.pt,color = cb!80] (2+\shift-2.1, -0.0) -- (2+\shift+0., -0.0); 
	\node at (1.5+\shift+0.65, -0., 3.1) {\tiny T1w-resembled};
    
    \draw [decorate,decoration={brace,amplitude=5pt,mirror,raise=6ex},line width=1.pt,color = tomato!60] (2+\shift+.3, -0.0) -- (2+\shift+2.4, -0.0);
	\node at (4+\shift+0.53, -0., 3.1) {\tiny T2w/FLAIR-resembled};

	\pgfmathsetmacro{\cubeza}{0.05*3}
	\pgfmathsetmacro{\cubezb}{0.07*3}
	\pgfmathsetmacro{\shift}{4.2}
	\foreach \i in {3.5}
	{
	\draw[black,fill=gray!30, line width = 0.02mm] (\i+\shift-0.1*3,0.*3,-0.1*3) -- ++(-\cubex,0,0) -- ++(0,-\cubey,0) -- ++(\cubex,0,0) -- cycle;
	\draw[black,fill=gray!35, line width = 0.02mm] (\i+\shift-0.1*3, 0.*3,-0.1*3) -- ++(0,0,-\cubezb) -- ++(0,-\cubey,0) -- ++(0,0,\cubezb) -- cycle;
	\draw[black,fill=gray!35, line width = 0.02mm] (\i+\shift-0.1*3,0.*3,-0.1*3) -- ++(-\cubex,0,0) -- ++(0,0,-\cubezb) -- ++(\cubex,0,0) -- cycle;
	
	\node at (\i+\shift-0.28*3, -0.18*3, -0.19*3) {\includegraphics[width=0.071\textwidth]{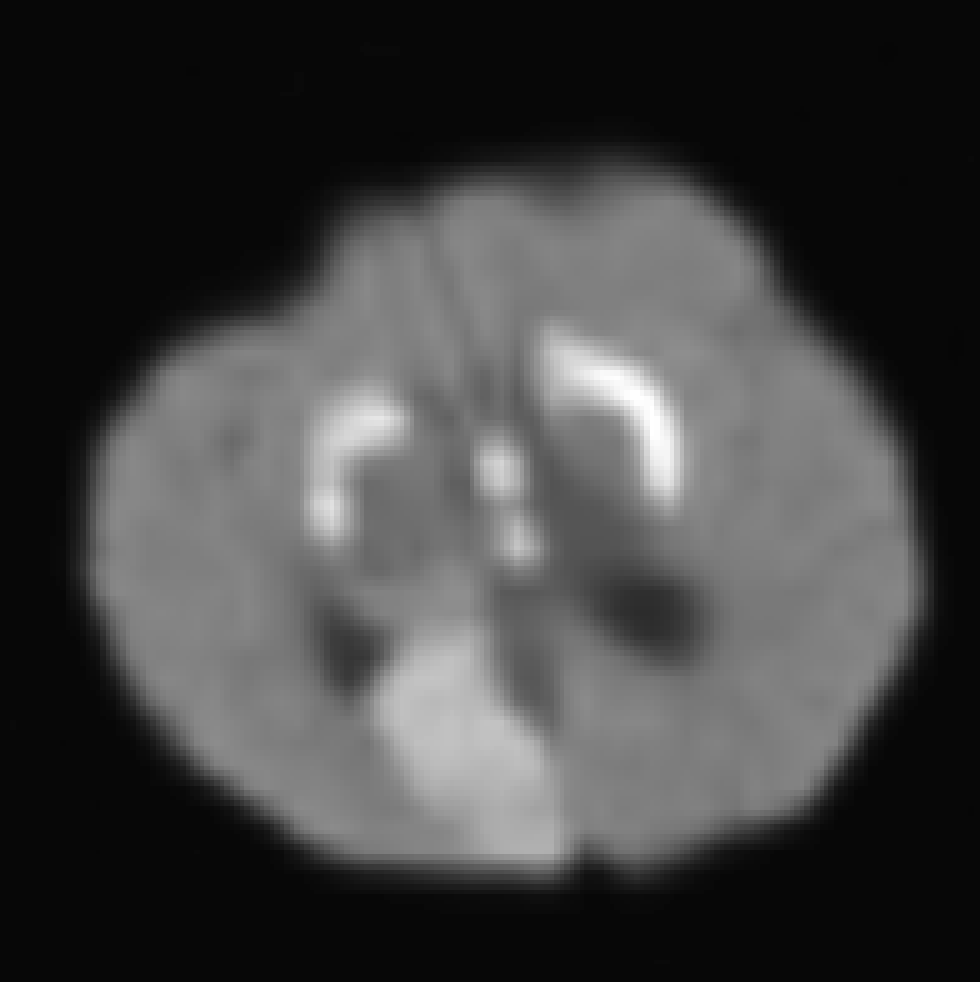}};
	\node at (\i+\shift+0.65, 0.23+1.3, 3.1) {\tiny $S_{N-1}$};
	}

	\pgfmathsetmacro{\cubeza}{0.07*3}
	\pgfmathsetmacro{\cubezb}{0.04*3}
	\pgfmathsetmacro{\shift}{4.2}
	\foreach \i in {4.5}
	{
	
	\draw[black,fill=gray!30, line width = 0.02mm] (\i+\shift-0.1*3,0.*3,-0.1*3) -- ++(-\cubex,0,0) -- ++(0,-\cubey,0) -- ++(\cubex,0,0) -- cycle;
	\draw[black,fill=gray!35, line width = 0.02mm] (\i+\shift-0.1*3, 0.*3,-0.1*3) -- ++(0,0,-\cubezb) -- ++(0,-\cubey,0) -- ++(0,0,\cubezb) -- cycle;
	\draw[black,fill=gray!35, line width = 0.02mm] (\i+\shift-0.1*3,0.*3,-0.1*3) -- ++(-\cubex,0,0) -- ++(0,0,-\cubezb) -- ++(\cubex,0,0) -- cycle;
	
	\node at (\i+\shift-0.28*3, -0.18*3, -0.19*3) {\includegraphics[width=0.071\textwidth]{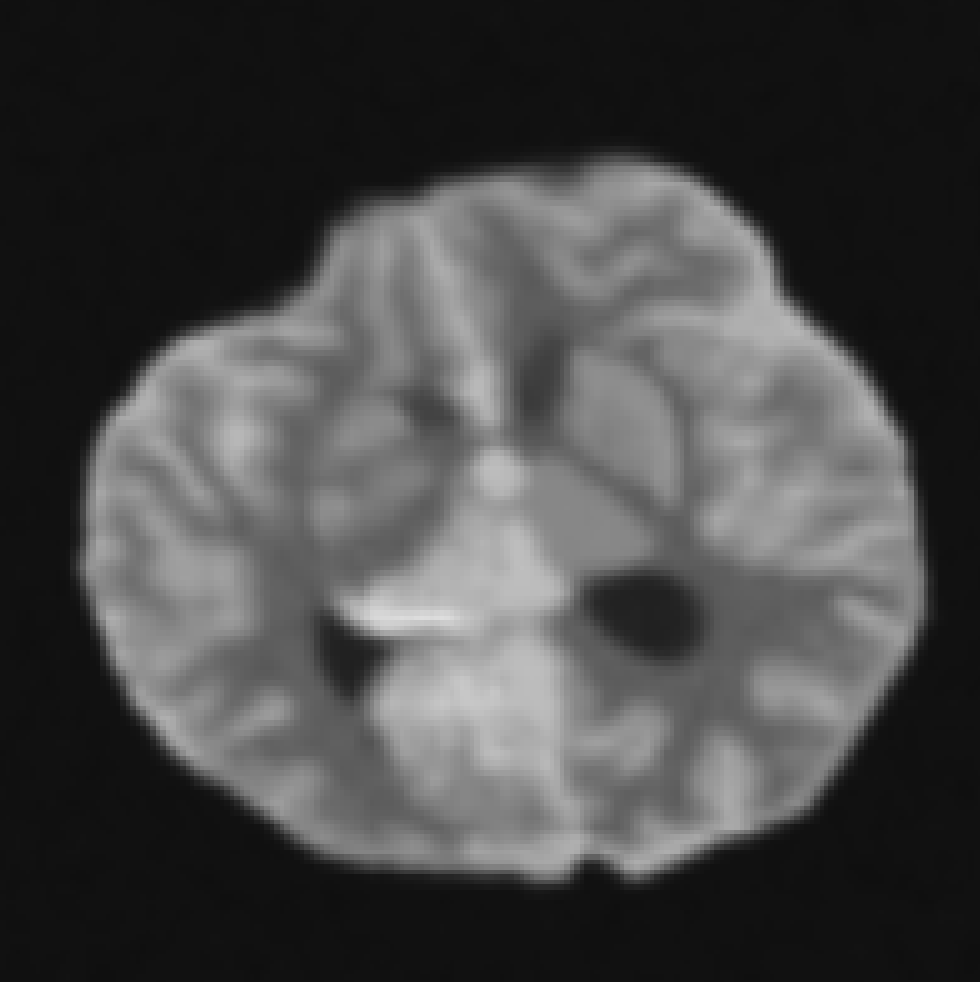}};
	\node at (\i+\shift+0.65, 0.23+1.3, 3.1) {\tiny $S_N$};
	}
	\end{tikzpicture}
	
}
\vspace{-0.35cm}
\caption{
\texttt{PEPSI}'s on-the-fly generator uses 3D anatomy labels ($L$) and anomaly probabilities ($P$) to generate training data with diverse deformations, contrasts, and corruptions -- enhanced by varying intensity profiles in pathological regions~(\cref{sec: generator}).}
\label{fig: augment}
\vspace{-0.3cm}
\end{figure}
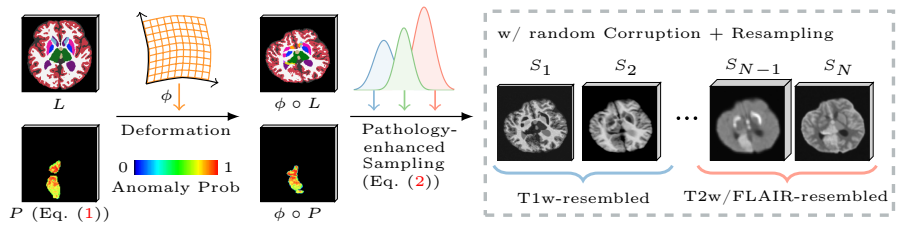

%% file: sec/method/trainer.tex
\vspace{-0.35cm}
\subsection{Representing across Contrasts, Pathologies, Datasets}
\label{sec: framework} 
\vspace{-0.2cm}

\input{sec/method/fw_train}

Here we present \texttt{PEPSI}'s training framework, which learns to emphasize anomalies and facilitates co-training across datasets with different types of pathology. 

\vspace{-0.5cm}
\subsubsection{Input:}
\label{sec: input}
We adopt the ``mild-to-severe'' intra-subject sampling strategy in~\cite{Liu2023BrainID}, which maximizes intra-subject variance to enhance feature robustness. Samples generated within a mini-batch are from the same subject, yet exhibit varying contrasts, corruptions, and pathology intensities, enriching the learning space~(\cref{fw: train}).

\vspace{-0.5cm}
\subsubsection{Dual Guidance Balancing Anatomy and Pathology:}
\label{sec: dual_supv}
We aim to obtain robust, contrast-agnostic feature representations that capture the distinctive anatomy of each subject while effectively distinguishing between pathology and normal tissue. MP-RAGE is the standard T1w MR contrast to delineate anatomical structures in research, but it is insufficient to differentiate many types of anomalies from normal tissue. FLAIR MRI, on the other hand, highlights areas of T2 prolongation as bright while suppressing cerebrospinal fluid (CSF), providing clear visibility of lesions in proximity to CSF~\cite{Hajnal1992FLAIR} -- but provides worse contrast than MP-RAGE in normal anatomy. \texttt{PEPSI} resorts to both MP-RAGE and FLAIR as learning targets, to concurrently capture normal anatomy and pathology. (\cref{fig: comp} compares the performance of dual-guidance and single-guidance.) 

As shown in \cref{fw: train}, the input mini-batch of intra-subject pathology-encoded samples, $\{ S_1, \, \dots, \, S_N \}$, are mapped to their corresponding feature space by a backbone ($\mathcal{F}$), $\{ \mathbf{F}_1, \, \dots, \, \mathbf{F}_N \}$. Two linear activation layers are followed to synthesize the anatomy and pathology images. The synthesis loss is obtained by collecting the reconstruction errors of all samples in the current mini-batch: 
\vspace{-0.25cm}
\begin{equation} 
\begin{aligned} 
    \mathcal{L}_{\text{synth}} &= \alpha \mathcal{L}_{I_\text{anat}} + \beta \mathcal{L}_{I_\text{pathol}} \qquad\qquad\qquad\qquad (\alpha,\,\beta \in \{0,\,1\})  \\
    &= \alpha \sum\nolimits_{i}^N ~ \vert \widetilde{I_i^{\text{anat}}} - I^{\text{anat}} \vert + \lambda \, \vert \nabla \widetilde{I_i^{\text{anat}}} - \nabla I^{\text{anat}} \vert \\  
    &\quad+ \beta \sum\nolimits_{i}^N ~ \vert \widetilde{I_i^{\text{pathol}}} - I^{\text{pathol}} \vert + \lambda \, \vert \nabla \widetilde{I_i^{\text{pathol}}} - \nabla I^{\text{pathol}} \vert,
    \label{eq: synth_loss}
\end{aligned}
\vspace{-0.22cm}
\end{equation}
where $\alpha,\,\beta$ denote the  availability of ground truth MP-RAGE ($I^{\text{anat}}$) and FLAIR ($I^{\text{pathol}}$), $\lambda \in R^+$ is the weight of reconstruction gradient loss~\cite{Liu2023BrainID}.


\input{sec/method/isles_case}
\vspace{-0.45cm}
\subsubsection{Implicit Pathology Supervision for Multi-pathology/dataset Training:}
\label{sec: impl_supv} 
\vspace*{-0.8cm}
\noindent Co-training across datasets broadens the model's exposure to various types of pathology, but also presents inherent challenges -- notably, difficulty to accurately synthesize \textit{abnormal} regions in the \textit{missing} modality, particularly for smaller datasets~(e.g., ``\texttt{PEPSI} (No-Seg)'' in \cref{fig: comp}). Direct supervision on pathology segmentations forces the model to pay more attention to anomalies, but could potentially result in conflicts during co-training due to the \textit{non-exhaustive} pathology annotations across datasets (e.g., ``\texttt{PEPSI} (Dir-Seg)'' in \cref{fig: comp}) -- The above figure shows a FLAIR image from ISLES~\cite{Hernandez2022ISLES} stroke dataset, despite the acquired FLAIR image clearly indicating WMH (circled in red), their gold-standard pathology segmentation \textit{only} provides/annotates areas of stroke lesions.

Here we propose an indirect pathology supervision approach. Specifically, for each output modality (i.e., MP-RAGE and FLAIR), we employ a ``third-party'', real-image-supervised pathology segmentation model as a reference, to encourage the pathology estimated from the predicted and ground truth images to align, without imposing strict supervision from the gold-standard pathology maps. As depicted in \cref{fw: train}, we pass all intra-subject training samples through the frozen, reference pathology segmentation models ($\mathcal{P}_{\text{anat}},\, \mathcal{P}_{\text{pathol}}$). The implicit pathology loss is computed based on the segmentation errors between the estimated pathology maps from the synthesized and ground truth images:
\vspace{-0.25cm}
\begin{equation} 
\begin{aligned} 
    \mathcal{L}_{\text{pathol}} &= \alpha \mathcal{L}_{S_\text{anat}} + \beta \mathcal{L}_{S_\text{pathol}} \qquad\qquad\qquad\qquad\qquad (\alpha,\,\beta \in \{0,\,1\})  \\
    &= \alpha \sum\nolimits_{i}^N ~ \mathcal{L}_{\text{seg}}( \widetilde{S_i^{\text{anat}}},\, S^{\text{anat}}) + \beta \sum\nolimits_{i}^N ~ \mathcal{L}_{\text{seg}}( \widetilde{S_i^{\text{pathol}}},\, S^{\text{pathol}})\,.
    \label{eq: pathol_loss}
\end{aligned}
\vspace{-0.22cm}
\end{equation}
$\mathcal{L}_{\text{seg}}$ is the segmentation loss consisting of soft dice and cross-entropy loss~\cite{Billot2021SynthSegSO}. Therefore, the overall training object writes $\mathcal{L} = \mathcal{L}_{\text{anat}} + \omega \mathcal{L}_{\text{pathol}},\, \omega \in \mathcal{R}^+ \label{eq: total_loss}$.

%% file: sec/method/fw_train.tex
\begin{figure}[t]
\centering 

\resizebox{\linewidth}{!}{
	\begin{tikzpicture}[lattice/.cd,spacing/.initial=4,superlattice
  period/.initial=12,amplitude/.initial=2]
		\tikzstyle{myarrows}=[line width=0.8mm,draw=blue!50,-triangle 45,postaction={draw, line width=0.05mm, shorten >=0.02mm, -}]
		\tikzstyle{mylines}=[line width=0.8mm]
  


    
 


	\pgfmathsetmacro{\cubex}{0.5*3}
	\pgfmathsetmacro{\cubey}{0.5*3}


	\pgfmathsetmacro{\shift}{0.5}
	\foreach \i/\cubez in {-5.9/0.12, -3.9/0.2, -0.5/0.3, 1.5/0.45}
	{
	\draw[black,fill=gray!35, line width = 0.02mm] (\shift+0.1*3,\i+0.15*3,0.1*3) -- ++(-\cubex,0,0) -- ++(0,-\cubey,0) -- ++(\cubex,0,0) -- cycle;
	\draw[black,fill=gray!35, line width = 0.02mm] (\shift+0.1*3,\i+0.15*3,0.1*3) -- ++(0,0,-\cubez) -- ++(0,-\cubey,0) -- ++(0,0,\cubez) -- cycle;
	\draw[black,fill=gray!35, line width = 0.02mm] (\shift+0.1*3,\i+0.15*3,0.1*3) -- ++(-\cubex,0,0) -- ++(0,0,-\cubez) -- ++(\cubex,0,0) -- cycle;
	}
 
	\node at (\shift+0.1*3, -5.9+0.15*3, 0.75*3) {\includegraphics[width=0.1220\textwidth]{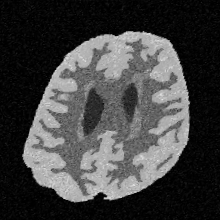}};
	\node at (\shift+0.1*3, -3.9+0.15*3, 0.75*3) {\includegraphics[width=0.1220\textwidth]{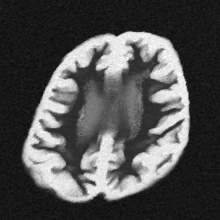}};
	\node at (\shift+0.1*3, -0.5+0.15*3, 0.75*3) {\includegraphics[width=0.1220\textwidth]{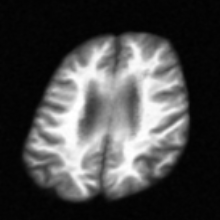}}; 
	\node at (\shift+0.1*3, 1.5+0.15*3, 0.75*3) {\includegraphics[width=0.1220\textwidth]{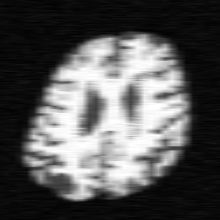}};
        \node at (0.+\shift+0.3-1.3, 1.5+0.15*3, 0.75*3) {\large$\mathbf{S}_N$};
        \node at (0.+\shift+0.3-1.3, -0.5+0.15*3, 0.75*3) {\large$\mathbf{S}_{N-1}$};
        \node at (0.+\shift+0.3-1.3, -3.9+0.15*3, 0.75*3) {\large$\mathbf{S}_2$};
        \node at (0.+\shift+0.3-1.3, -5.9+0.15*3, 0.75*3) {\large$\mathbf{S}_1$};
 
    \draw[black,fill=black] (\shift+0.3, -2.+0.15*3, 0.75*3) circle (0.8pt);
    \draw[black,fill=black] (\shift+0.3, -2.2+0.15*3, 0.75*3) circle (0.8pt);
    \draw[black,fill=black] (\shift+0.3, -2.4+0.15*3, 0.75*3) circle (0.8pt);

    \node at (\shift+0.3, -7.2+0.15*3, 0.75*3) {\large Subject-$idx$~(\cref{sec: generator})};

	\pgfmathsetmacro{\shift}{0.95}
        \pgfmathsetmacro{\dx}{-6.}
        \pgfmathsetmacro{\dy}{-2.3}
	\draw[dashed, color = cb, line width=0.4mm] (4+\dx, -5.+\dy) -- (7.2+\dx, -5.+\dy) -- (7.2+\dx, 4.5+\dy) -- (4+\dx, 4.5+\dy) -- (4+\dx, -5.+\dy);

    \node at (\shift+3.57, -5.2+0.15*3, 0.75*3) {\large Modality availability};
    
	\pgfmathsetmacro{\shift}{0.8}
    \node at (\shift+2.2, -5.7+0.15*3, 0.75*3) {\large\textcolor{applegreen}{\cmark}};
    \node at (\shift+3.246, -5.7+0.15*3, 0.75*3) {\large: available};
    
    \node at (\shift+2.2, -6.2+0.15*3, 0.75*3) {\large\textcolor{red}{\xmark}};
    \node at (\shift+3.46, -6.2+0.15*3, 0.75*3) {\large: unavailable};

	\pgfmathsetmacro{\shift}{1.}
        \pgfmathsetmacro{\dx}{-2.25}
        \pgfmathsetmacro{\dy}{-2.}
	\draw[dashed, color = cb, line width=0.4mm] (4-0.05+\dx, -5.+\dy) -- (7.9-0.05+\dx, -5.+\dy) -- (7.9-0.05+\dx, -3.2+\dy) -- (4-0.05+\dx, -3.2+\dy) -- (4-0.05+\dx, -5.+\dy);

    \draw [decorate,decoration={brace,amplitude=5pt,raise=6ex},line width=2.pt,color = cb] (0.6, 2) -- (0.6, -7.1);

    
	\pgfmathsetmacro{\shift}{0.5}
    \pgfmathsetmacro{\sx}{2.1}
    \pgfmathsetmacro{\sy}{5.4}
    \pgfmathsetmacro{\dx}{\sx+0}
    \pgfmathsetmacro{\dy}{\sy+0.2} 
    \pgfmathsetmacro{\ddy}{-3.5} 
    \pgfmathsetmacro{\dxt}{\dx+0.5} 
    
    \node at (1.2+\sx, -6.2+\dy+\ddy){\large $\mathcal{F}$: backbone};

    \pgfmathsetmacro{\dy}{\dy+\ddy} 
    \networkLayer{2}{0.1}{5.1-\dy+\dxt}{13-2.6*\dy}{color=myblue!80}{}
    \networkLayer{1.6}{0.2}{5.2-\dy+\dxt}{13-2.6*\dy}{color=myblue!60}{}
    \networkLayer{1.2}{0.4}{5.4-\dy+\dxt}{13-2.6*\dy}{color=myblue!40}{}
    \networkLayer{0.8}{0.8}{5.8-\dy+\dxt}{13-2.6*\dy}{color=myblue!20}{}
    
    \pgfmathsetmacro{\dxs}{\dx+0}
    \pgfmathsetmacro{\dys}{\dy+0} 
    \networkLayer{0.8}{0.8}{7.2-\dys+\dxs}{13-2.6*\dys}{color=matcha!20}{}
    \networkLayer{1.2}{0.4}{7.6-\dys+\dxs}{13-2.6*\dys}{color=matcha!40}{}
    \networkLayer{1.6}{0.2}{8.-\dys+\dxs}{13-2.6*\dys}{color=matcha!60}{}
    \networkLayer{2}{0.1}{8.3-\dys+\dxs}{13-2.6*\dys}{color=matcha!80}{}

 
    \draw [decorate,decoration={brace,amplitude=5pt,mirror,raise=6ex},line width=2.pt,color = matcha!120] (6.72, 2) -- (6.72, -7.1);

	\pgfmathsetmacro{\cubez}{0.12}  
	\pgfmathsetmacro{\shift}{1.2}
 
	\foreach \i/\j in {7.2/-5.9, 7.2/-3.9, 7.2/-0.5, 7.2/1.5,  9.2/-5.9, 9.2/-3.9, 9.2/-0.5, 9.2/1.5,  12.6/-5.9, 12.6/-3.9, 12.6/-0.5, 12.6/1.5}
	{
	\draw[black,fill=gray!35, line width = 0.02mm] (\i+\shift+0.1*3,\j+0.15*3,0.1*3) -- ++(-\cubex,0,0) -- ++(0,-\cubey,0) -- ++(\cubex,0,0) -- cycle;
	\draw[black,fill=gray!35, line width = 0.02mm] (\i+\shift+0.1*3,\j+0.15*3,0.1*3) -- ++(0,0,-\cubez) -- ++(0,-\cubey,0) -- ++(0,0,\cubez) -- cycle;
	\draw[black,fill=gray!35, line width = 0.02mm] (\i+\shift+0.1*3,\j+0.15*3,0.1*3) -- ++(-\cubex,0,0) -- ++(0,0,-\cubez) -- ++(\cubex,0,0) -- cycle;
 
        \draw[black,fill=black] (\i+\shift+0.3, -2.+0.15*3, 0.75*3) circle (0.8pt);
        \draw[black,fill=black] (\i+\shift+0.3, -2.2+0.15*3, 0.75*3) circle (0.8pt);
        \draw[black,fill=black] (\i+\shift+0.3, -2.4+0.15*3, 0.75*3) circle (0.8pt);
        
        \draw[black,fill=black] (10.9-0.2+\shift+0.3, \j+0.15*3, 0.75*3) circle (0.8pt);
        \draw[black,fill=black] (10.9+\shift+0.3, \j+0.15*3, 0.75*3) circle (0.8pt);
        \draw[black,fill=black] (10.9+0.2+\shift+0.3, \j+0.15*3, 0.75*3) circle (0.8pt);
        
        \draw[black,fill=black] (10.9-0.2+\shift+0.3, -2.+0.15*3, 0.75*3) circle (0.8pt);
        \draw[black,fill=black] (10.9+\shift+0.3, -2.2+0.15*3, 0.75*3) circle (0.8pt);
        \draw[black,fill=black] (10.9+0.2+\shift+0.3, -2.4+0.15*3, 0.75*3) circle (0.8pt);
	}

        \pgfmathsetmacro{\dx}{2.9}
        \foreach \dy in {-2.2, -0.2, 3.2, 5.2}
        {
	\draw[dashed, color = matcha!120, line width=0.4mm] (4+\dx, -5.+\dy) -- (11.3+\dx, -5.+\dy) -- (11.3+\dx, -3.2+\dy) -- (4+\dx, -3.2+\dy) -- (4+\dx, -5.+\dy); 
        }
        
	\node at (7.2+\shift+0.1*3, 1.5+0.15*3, 0.75*3) {\includegraphics[width=0.1220\textwidth]{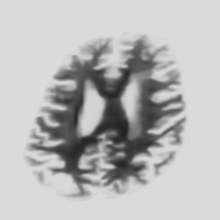}};
	\node at (7.2+\shift+0.1*3, -0.5+0.15*3, 0.75*3) {\includegraphics[width=0.1220\textwidth]{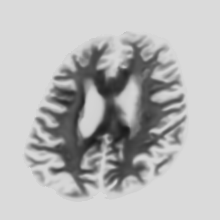}};
	\node at (7.2+\shift+0.1*3, -3.9+0.15*3, 0.75*3) {\includegraphics[width=0.1220\textwidth]{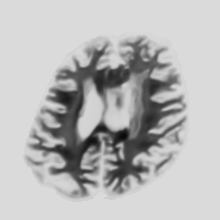}};
	\node at (7.2+\shift+0.1*3, -5.9+0.15*3, 0.75*3) {\includegraphics[width=0.1220\textwidth]{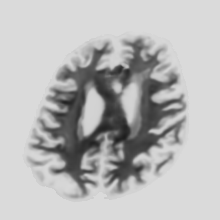}};
 
	\node at (9.2+\shift+0.1*3, 1.5+0.15*3, 0.75*3) {\includegraphics[width=0.1220\textwidth]{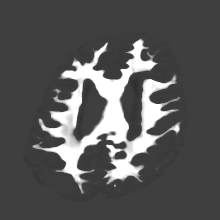}};
	\node at (9.2+\shift+0.1*3, -0.5+0.15*3, 0.75*3) {\includegraphics[width=0.1220\textwidth]{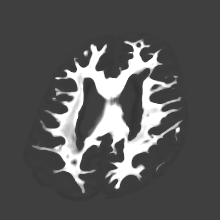}};
	\node at (9.2+\shift+0.1*3, -3.9+0.15*3, 0.75*3) {\includegraphics[width=0.1220\textwidth]{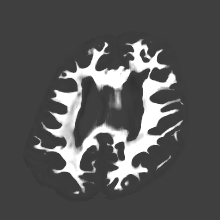}};
	\node at (9.2+\shift+0.1*3, -5.9+0.15*3, 0.75*3) {\includegraphics[width=0.1220\textwidth]{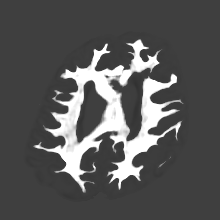}};
 
	\node at (12.6+\shift+0.1*3, 1.5+0.15*3, 0.75*3) {\includegraphics[width=0.1220\textwidth]{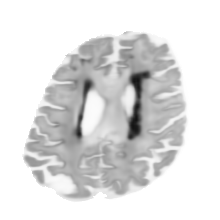}};
	\node at (12.6+\shift+0.1*3, -0.5+0.15*3, 0.75*3) {\includegraphics[width=0.1220\textwidth]{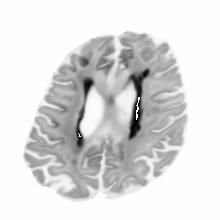}};
	\node at (12.6+\shift+0.1*3, -3.9+0.15*3, 0.75*3) {\includegraphics[width=0.1220\textwidth]{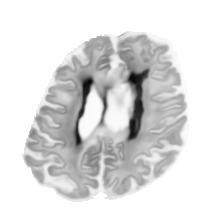}};
	\node at (12.6+\shift+0.1*3, -5.9+0.15*3, 0.75*3) {\includegraphics[width=0.1220\textwidth]{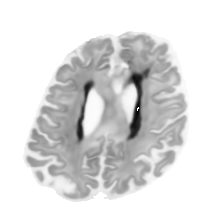}};

        \node at (7.2+\shift+0.36-1.5, 1.5+0.15*3, 0.75*3) {\large$\mathbf{F}_{N}$};
        \node at (7.2+\shift+0.36-1.5, -0.5+0.15*3, 0.75*3) {\large$\mathbf{F}_{N-1}$};
        \node at (7.2+\shift+0.36-1.5, -3.9+0.15*3, 0.75*3) {\large$\mathbf{F}_2$};
        \node at (7.2+\shift+0.36-1.5, -5.9+0.15*3, 0.75*3) {\large$\mathbf{F}_1$};

	\pgfmathsetmacro{\shift}{1.6}
 
        \draw [mylines, color = matcha](11.8+\shift+0.3+2.1, -5.9+0.15*3, 0.75*3) -- (11.4+\shift+2.3, -5.9+0.15*3, 0.75*3);  
        
        \draw [mylines, color = matcha](11.8+\shift+0.3+2.1, -3.9+0.15*3, 0.75*3) -- (11.4+\shift+2.3, -3.9+0.15*3, 0.75*3);  
        
        \draw [mylines, color = matcha](11.8+\shift+0.3+2.1, -0.5+0.15*3, 0.75*3) -- (11.4+\shift+2.3, -0.5+0.15*3, 0.75*3);  
        
        \draw [mylines, color = matcha](11.8+\shift+0.3+2.1, 1.5+0.15*3, 0.75*3) -- (11.4+\shift+2.3, 1.5+0.15*3, 0.75*3);  
        
        \draw [mylines, color = matcha](11.8+\shift+0.3+2.1, -5.9+0.15*3, 0.75*3) -- (11.8+\shift+0.3+2.1, 1.5+0.15*3, 0.75*3); 
        
        \draw [mylines, color = matcha](11.8+\shift+0.3+2.1, -2.2+0.15*3, 0.75*3) -- (12.2+\shift+0.3+2.3, -2.2+0.15*3, 0.75*3);
        

	\pgfmathsetmacro{\shift}{0.6}
 
        \draw [mylines, color = matcha](13.2+\shift+0.3+2.3, -1+0.15*3, 0.75*3) -- (13.2+\shift+0.3+2.3, -3.4+0.15*3, 0.75*3);  
        
        \draw [myarrows, color = matcha](13.2+\shift+0.3+2.3, -1+0.15*3, 0.75*3) -- (14.+\shift+0.3+2.3, -1+0.15*3, 0.75*3); 
        
        \draw [myarrows, color = matcha](13.2+\shift+0.3+2.3, -3.4+0.15*3, 0.75*3) -- (14.+\shift+0.3+2.3, -3.4+0.15*3, 0.75*3);

	\pgfmathsetmacro{\shift}{1.8}
    
    \foreach \i/\j in {17.1/-1., 17.1/-1., 17.1/1.6, 21.1/1.6, 21.1/-1., 17.1/-3.4, 21.1/-3.4, 17.1/-5.9, 21.1/-5.9}{
    \draw[black,fill=gray!35, line width = 0.02mm] (\i+\shift+0.1*3,\j+0.15*3,0.1*3) -- ++(-\cubex,0,0) -- ++(0,-\cubey,0) -- ++(\cubex,0,0) -- cycle;
    \draw[black,fill=gray!35, line width = 0.02mm] (\i+\shift+0.1*3,\j+0.15*3,0.1*3) -- ++(0,0,-\cubez) -- ++(0,-\cubey,0) -- ++(0,0,\cubez) -- cycle;
    \draw[black,fill=gray!35, line width = 0.02mm] (\i+\shift+0.1*3,\j+0.15*3,0.1*3) -- ++(-\cubex,0,0) -- ++(0,0,-\cubez) -- ++(\cubex,0,0) -- cycle;
    }

    \pgfmathsetmacro{\i}{17.1}
    \pgfmathsetmacro{\j}{-1.}
    \node at (\i+\shift+0.1*3, \j+0.15*3, 0.75*3) {\includegraphics[width=0.1220\textwidth]{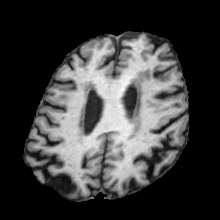}};
    \node at (\i-1.4+\shift+0.1*3, \j+0.15*3, 0.75*3) {\large$\widetilde{I_i^{\text{anat}}}$};

    \pgfmathsetmacro{\i}{17.1}
    \pgfmathsetmacro{\j}{1.6}
    \node at (\i+\shift+0.1*3, \j+0.15*3, 0.75*3) {\includegraphics[width=0.1220\textwidth]{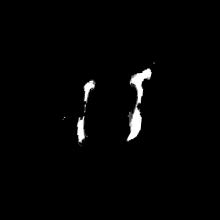}};
    \node at (\i-1.42+\shift+0.1*3, \j+0.15*3, 0.75*3) {\large$\widetilde{S_i^{\text{anat}}}$};

    \pgfmathsetmacro{\i}{21.1}
    \pgfmathsetmacro{\j}{-1.}  
    \node at (\i+\shift+0.1*3, \j+0.15*3, 0.75*3) {\includegraphics[width=0.1220\textwidth]{fig/fw/gt_t1.png}};
    \node at (\i+1.4+\shift+0.1*3, \j+0.15*3, 0.75*3) {\large${I^{\text{anat}}}$}; 
    \node at (\i+1.4+\shift+0.1*3, \j+0.15*3-0.5, 0.75*3) {\large ({\color{applegreen}\cmark}/{\color{red}\xmark})}; 
    
    \node at (\i-2+\shift+0.1*3, \j-0.4+0.15*3, 0.75*3) {\large\cref{eq: synth_loss}};

    \pgfmathsetmacro{\i}{21.1}
    \pgfmathsetmacro{\j}{1.6}  
    \node at (\i+\shift+0.1*3, \j+0.15*3, 0.75*3) {\includegraphics[width=0.1220\textwidth]{fig/fw/gt_pathol_134.png}};
    \node at (\i+1.42+\shift+0.1*3, \j+0.15*3, 0.75*3) {\large${S^{\text{anat}}}$}; 
    
    \node at (\i-2+\shift+0.1*3, \j+0.4+0.15*3, 0.75*3) {\large\cref{eq: pathol_loss}};

    \pgfmathsetmacro{\i}{16.1}
    \pgfmathsetmacro{\j}{-1}
    
    \draw [<-, color = flatgrey, dotted, line width = 0.7mm](\i-0.4+\shift+0.3+2.3, \j+0.15*3-0.4, 0.75*3) -- (\i+0.+\shift+0.3+2.3, \j+0.15*3-0.4, 0.75*3);
    \draw [->, color = flatgrey, dotted, line width = 0.7mm](\i+1.7-0.3+\shift+0.3+2.3, \j+0.15*3-0.4, 0.75*3) -- (\i+1.7+0.1+\shift+0.3+2.3, \j+0.15*3-0.4, 0.75*3);


    \pgfmathsetmacro{\j}{-1}
    \draw [myarrows, color = tomato!60, dotted](\i+\shift+0.3+2.1, \j+0.15*3+0.4, 0.75*3) -- (\i+0.5+\shift+0.3+2.3, \j+0.15*3+0.4, 0.75*3) -- (\i+0.5+\shift+0.3+2.3, \j+1.8+0.15*3+0.4, 0.75*3) -- (\i-0.2+\shift+0.3+2.1, \j+1.8+0.15*3+0.4, 0.75*3);
    \draw [myarrows, color = tomato!60, dotted](\i+1.6+\shift+0.3+2.3, \j+0.15*3+0.4, 0.75*3) -- (\i+1.+\shift+0.3+2.3, \j+0.15*3+0.4, 0.75*3) -- (\i+1.+\shift+0.3+2.3, \j+1.8+0.15*3+0.4, 0.75*3) -- (\i+1.8+\shift+0.3+2.3, \j+1.8+0.15*3+0.4, 0.75*3);

    \pgfmathsetmacro{\x}{19.63} 
    \pgfmathsetmacro{\y}{-.5} 
    \draw[rounded corners, draw=tomato!60, fill = tomato!20, line width=0.6mm] (\x, \y) rectangle (\x+1.5, \y+0.8) {};
    \node at (\x+0.75,\y+0.4) {\large $\mathcal{P}_{\text{anat}}$};

    \pgfmathsetmacro{\j}{1.6} 
    \draw [<-, color = flatgrey, dotted, line width = 0.7mm](\i-0.4+\shift+0.3+2.3, \j+0.15*3+0.4, 0.75*3) -- (\i+0.+\shift+0.3+2.3, \j+0.15*3+0.4, 0.75*3);
    \draw [->, color = flatgrey, dotted, line width = 0.7mm](\i+1.7-0.3+\shift+0.3+2.3, \j+0.15*3+0.4, 0.75*3) -- (\i+1.7+0.1+\shift+0.3+2.3, \j+0.15*3+0.4, 0.75*3);

    \pgfmathsetmacro{\j}{-3.4}
    
    \draw [myarrows, color = tomato!60, dotted](\i+\shift+0.3+2.1, \j+0.15*3-0.4, 0.75*3) -- (\i+0.5+\shift+0.3+2.3, \j+0.15*3-0.4, 0.75*3) -- (\i+0.5+\shift+0.3+2.3, \j-1.8+0.15*3-0.4, 0.75*3) -- (\i-0.2+\shift+0.3+2.1, \j-1.8+0.15*3-0.4, 0.75*3);
    \draw [myarrows, color = tomato!60, dotted](\i+1.6+\shift+0.3+2.3, \j+0.15*3-0.4, 0.75*3) -- (\i+1.+\shift+0.3+2.3, \j+0.15*3-0.4, 0.75*3) -- (\i+1.+\shift+0.3+2.3, \j-1.8+0.15*3-0.4, 0.75*3) -- (\i+1.8+\shift+0.3+2.3, \j-1.8+0.15*3-0.4, 0.75*3);

    \draw [<-, color = flatgrey, dotted, line width = 0.7mm](\i-0.4+\shift+0.3+2.3, \j+0.15*3+0.4, 0.75*3) -- (\i+0.+\shift+0.3+2.3, \j+0.15*3+0.4, 0.75*3);
    \draw [->, color = flatgrey, dotted, line width = 0.7mm](\i+1.7-0.3+\shift+0.3+2.3, \j+0.15*3+0.4, 0.75*3) -- (\i+1.7+0.1+\shift+0.3+2.3, \j+0.15*3+0.4, 0.75*3);

    \pgfmathsetmacro{\j}{-6.7} 
    \draw [<-, color = flatgrey, dotted, line width = 0.7mm](\i-0.4+\shift+0.3+2.3, \j+0.15*3+0.4, 0.75*3) -- (\i+0.+\shift+0.3+2.3, \j+0.15*3+0.4, 0.75*3);
    \draw [->, color = flatgrey, dotted, line width = 0.7mm](\i+1.7-0.3+\shift+0.3+2.3, \j+0.15*3+0.4, 0.75*3) -- (\i+1.7+0.1+\shift+0.3+2.3, \j+0.15*3+0.4, 0.75*3);
    
    \pgfmathsetmacro{\x}{19.63} 
    \pgfmathsetmacro{\y}{-5.5} 
    \draw[rounded corners, draw=tomato!60, fill = tomato!20, line width=0.6mm] (\x, \y) rectangle (\x+1.5, \y+0.8) {};
    \node at (\x+0.75,\y+0.4) {\large$\mathcal{P}_{\text{pathol}}$};

    \pgfmathsetmacro{\i}{17.1}
    \pgfmathsetmacro{\j}{-3.4} 
    \node at (\i+\shift+0.1*3, \j+0.15*3, 0.75*3) {\includegraphics[width=0.1220\textwidth]{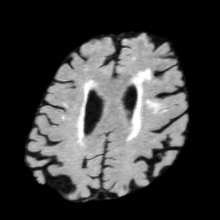}};
    \node at (\i-1.4+\shift+0.1*3, \j+0.15*3, 0.75*3) {\large$\widetilde{I_i^{\text{pathol}}}$};
    
    \node at (\i+2+\shift+0.1*3, \j+0.4+0.15*3, 0.75*3) {\large\cref{eq: synth_loss}};
    
    \pgfmathsetmacro{\i}{17.1}
    \pgfmathsetmacro{\j}{-5.9} 
    \node at (\i+\shift+0.1*3, \j+0.15*3, 0.75*3) {\includegraphics[width=0.1220\textwidth]{fig/fw/gt_pathol_134.png}};
    \node at (\i-1.43+\shift+0.1*3, \j+0.15*3, 0.75*3) {\large$\widetilde{S_i^{\text{pathol}}}$};
    
    \node at (\i+2+\shift+0.1*3, \j-0.4+0.15*3, 0.75*3) {\large\cref{eq: pathol_loss}};

    \pgfmathsetmacro{\i}{21.1}
    \pgfmathsetmacro{\j}{-3.4} 
    \node at (\i+\shift+0.1*3, \j+0.15*3, 0.75*3) {\includegraphics[width=0.1220\textwidth]{fig/fw/gt_flair.png}};
    \node at (\i+1.4+\shift+0.1*3, \j+0.15*3, 0.75*3) {\large${I^{\text{pathol}}}$};
    \node at (\i+1.4+\shift+0.1*3, \j+0.15*3-0.5, 0.75*3) {\large ({\color{applegreen}\cmark}/{\color{red}\xmark})};
    
    \pgfmathsetmacro{\i}{21.1}
    \pgfmathsetmacro{\j}{-5.9} 
    \node at (\i+\shift+0.1*3, \j+0.15*3, 0.75*3) {\includegraphics[width=0.1220\textwidth]{fig/fw/gt_pathol_134.png}};
    \node at (\i+1.43+\shift+0.1*3, \j+0.15*3, 0.75*3) {\large${S^{\text{pathol}}}$};


    \pgfmathsetmacro{\dx}{6.9}
    \pgfmathsetmacro{\dy}{0}
    \draw [myarrows, color = matcha!120](-0.5+\dx, -7.35+\dy) -- (8+\dx, -7.35+\dy);  
    \node[anchor=north] at (3.7+\dx, -7.4+\dy) {\textcolor{matcha!150}{\large\textbf{Feature Channel}}};

       
	\end{tikzpicture}
	}  
	\vspace{-0.6cm}
\caption{\texttt{PEPSI}'s pathology-enhanced, contrast-agnostic training overview~(\cref{sec: framework}).}
	 \label{fw: train}
\end{figure}

%% file: sec/method/isles_case.tex
\begin{figure}[h]
    \vspace*{-0.56cm}
    \resizebox{\textwidth}{!}{
    \begin{tikzpicture}[lattice/.cd,spacing/.initial=4,superlattice
  period/.initial=12,amplitude/.initial=2]
		\tikzstyle{myarrows}=[line width=0.6mm,draw=blue!50,-triangle 45,postaction={draw, line width=0.05mm, shorten >=0.02mm, -}]
		\tikzstyle{mylines}=[line width=0.8mm]
    \node at (0, 0) {\includegraphics[width=0.2\textwidth]{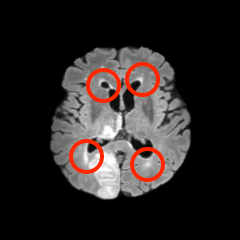}};
    
     \node at (2.6, 0) {\includegraphics[width=0.2\textwidth]{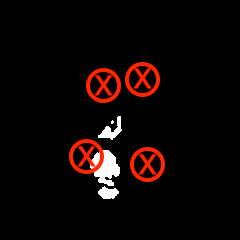}};

         \node at (-4.5+0.5, 0) {ISLES~\cite{Hernandez2022ISLES} FLAIR};
        \draw [myarrows, color = flatgrey](-3+0.5, 0) -- (-1.5, 0); 
        \draw [myarrows, color = flatgrey](6-0.4-0.5, 0) -- (4.5-0.4, 0); 
         \node at (7.5, 0) {Gold-standard pathology map};
         \node at (7.5, -0.5) {(Stroke annotations \textit{only})};

    \end{tikzpicture}   
    }
\end{figure}

%% file: sec/exp/main.tex
\input{sec/exp/tab_synth}

\input{sec/exp/fig_synth}

\vspace{-0.2cm}
\section{Experiments}
\label{sec: exp}
\vspace{-0.22cm}

We demonstrate the effectiveness of \texttt{PEPSI} from two perspectives: \textit{(i) Image synthesis} --- estimating both anatomy and pathology images, with potentially missing modalities~(\cref{exp: pretrain}); \textit{(ii) Pathology segmentation} --- fine-tuning \texttt{PEPSI} on individual datasets for segmenting a specific type of pathology~(\cref{exp: downstream}).

\input{sec/exp/setup}
\input{sec/exp/synthesis}


\input{sec/exp/comp_curve}

\input{sec/exp/tab_seg}

\input{sec/exp/fig_seg}
\input{sec/exp/segment}


%% file: sec/exp/tab_synth.tex
\begin{table}[t]
    \caption{Quantitative comparisons in anatomy and pathology image synthesis among \texttt{PEPSI}, its variants, and the state-of-the-art contrast-agnostic synthesis models. The proposed \texttt{PEPSI} \textit{(i)} outperforms all the other models, especially on single-modality datasets; and \textit{(ii)} preserves its high performance even for cross-modality synthesis.
    } 
    \vspace*{-0.2cm}
    \label{tab: synth}
\resizebox{\linewidth}{!}{
\centering 
    \begin{tabular}{cccrccccccc}
       \toprule \\[-3ex] 
      \multicolumn{1}{c}{\multirow{2}{*}{\thead{\footnotesize\textbf{Dataset}\\{\footnotesize{~(\# of train/test)~}}}}} & \multicolumn{2}{c}{\multirow{1}{*}{\footnotesize\textbf{MR Contrast}}} & \multicolumn{1}{c}{\multirow{2}{*}{\footnotesize\textbf{Metric}}} & \multicolumn{1}{c}{\multirow{2}{*}{\thead{\footnotesize\texttt{~SynthSR~}\\{~\footnotesize{\cite{Iglesias2023SynthSRAP}}~}}}} & \multicolumn{1}{c}{\multirow{2}{*}{\thead{\footnotesize\texttt{~Brain-ID~}\\{~\footnotesize{\cite{Liu2023BrainID}}~}}}} & \multicolumn{1}{c}{\multirow{2}{*}{\thead{\footnotesize\texttt{PEPSI}\\{\footnotesize{~(SG-Anat)~}}}}} & \multicolumn{1}{c}{\multirow{2}{*}{\thead{\footnotesize\texttt{PEPSI}\\{\footnotesize{~(SG-Pathol)~}}}}}  & \multicolumn{1}{c}{\multirow{2}{*}{\thead{\footnotesize\texttt{PEPSI}\\{\footnotesize{~(No-Seg)~}}}}} & \multicolumn{1}{c}{\multirow{2}{*}{\thead{\footnotesize\texttt{PEPSI}\\{\footnotesize{~(Dir-Seg)~}}}}} & \multicolumn{1}{c}{\multirow{2}{*}{\thead{\footnotesize\color{black}\texttt{PEPSI}\\{\footnotesize{~(\color{black}{Proposed})~}}}}} \\ [-0.5ex]
        \cmidrule(lr){2-3}
          & {\footnotesize{Input}} & {\footnotesize{Output}} & & & & & & & &   \\ [-0.2ex]
     \midrule\\[-3ex]

        \multicolumn{1}{c}{\multirow{3}{*}{\thead{\footnotesize{ATLAS~\cite{Liew2017ATLAS}}\\(590/65)}}} & \multicolumn{1}{c}{\multirow{3}{*}{{ \footnotesize{~T1w~}}}} & \multicolumn{1}{c}{\multirow{3}{*}{{ \footnotesize{~T1w~}}}} & ~{\footnotesize {\texttt{L1} ($\downarrow$)}}~ & 0.067 & 0.65 & 0.69 & - & 0.052 & 0.074 & \textbf{0.036} \\ 
        & & & ~{\footnotesize {\texttt{PSNR} ($\uparrow$)}}~ & 16.90 & 17.91 & 16.54 & - & 18.46 & 16.01 & \textbf{21.69}\\ 
        & & & ~{\footnotesize {\texttt{SSIM} ($\uparrow$)}}~ & 0.804 & 0.833 & 0.845 & - & 0.861 & 0.831 & \textbf{0.897} \\ 
         \hline 

     \multicolumn{1}{c}{\multirow{3.2}{*}{\thead{\footnotesize{ISLES~\cite{Hernandez2022ISLES}}\\(137/15)}}} & \multicolumn{1}{c}{\multirow{3}{*}{\footnotesize{~FLAIR~}}} & \multicolumn{1}{c}{\multirow{3}{*}{\footnotesize{~FLAIR~}}}  & ~{\footnotesize {\texttt{L1} ($\downarrow$)}}~ & - & - & - & 0.022 & 0.018 & 0.021 & \textbf{0.016} \\ 
        &  &  & ~{\footnotesize {\texttt{PSNR} ($\uparrow$)}}~ & - & - & - & 23.87 & 25.34 & 24.02 & \textbf{26.03} \\ 
        &  &  & ~{\footnotesize {\texttt{SSIM} ($\uparrow$)}}~ & - & - & - & 0.962 & 0.942 & 0.926 & \textbf{0.969} \\ 
         \hline 
     
     \multicolumn{1}{c}{\multirow{12.6}{*}{\thead{\footnotesize{~ADNI3~\cite{Weiner2017TheAD}}\\(298/33)~}}} & \multicolumn{1}{c}{\multirow{6}{*}{\footnotesize{~T1w~}}} & \multicolumn{1}{c}{\multirow{3}{*}{\footnotesize{~T1w~}}} & ~{\footnotesize {\texttt{L1} ($\downarrow$)}}~ & 0.023 & 0.021 & 0.025 & - & 0.022 & 0.022 & \textbf{0.020} \\  
        & & & ~{\footnotesize {\texttt{PSNR} ($\uparrow$)}}~ & 23.51 & 24.42 & 24.44 & - & 24.01 & 23.37 & \textbf{26.67} \\  
        & & & ~{\footnotesize {\texttt{SSIM} ($\uparrow$)}}~ & 0.901 & 0.899 & 0.930 & - & 0.932 & 0.931 & \textbf{0.935} \\  [-0.5ex] 
        
        \cmidrule(lr){3-11}
     
         &  & \multicolumn{1}{c}{\multirow{3}{*}{\footnotesize{~FLAIR~}}} & ~{\footnotesize {\texttt{L1} ($\downarrow$)}}~ & - & - & - & 0.043 & 0.392 & 0.396 &\textbf{0.036} \\ 
        & & & ~{\footnotesize {\texttt{PSNR} ($\uparrow$)}}~ & - & - & - & 18.87 & 19.64 & 19.58 & \textbf{21.40} \\ 
        & & & ~{\footnotesize {\texttt{SSIM} ($\uparrow$)}}~ & - & - & - & 0.900 & 0.901 & 0.894 &\textbf{0.911}  \\  [-0.5ex]

        \cmidrule(lr){2-11}
        
        & \multicolumn{1}{c}{\multirow{6}{*}{\footnotesize{~FLAIR~}}} & \multicolumn{1}{c}{\multirow{3}{*}{\footnotesize{~T1w~}}} &  ~{\footnotesize {\texttt{L1} ($\downarrow$)}}~ & 0.027 & 0.026 & 0.027 & - & 0.027 & 0.029 & \textbf{0.023}\\ 
         & & & ~{\footnotesize {\texttt{PSNR} ($\uparrow$)}}~ & 23.25 & 23.74 & 23.96 & - & 23.50 & 23.61 & \textbf{25.62} \\ 
         & & & ~{\footnotesize {\texttt{SSIM} ($\uparrow$)}}~ & 0.906 & 0.879 & 0.916 & - & 0.919 & 0.915 & \textbf{0.929} \\ [-0.5ex] 
        
        \cmidrule(lr){3-11}
        
         & & \multicolumn{1}{c}{\multirow{3}{*}{\footnotesize{~FLAIR~}}} &  ~{\footnotesize {\texttt{L1} ($\downarrow$)}}~ & - & - & - & 0.044 & 0.0396 & 0.041 & \textbf{0.034}  \\ 
         & & & ~{\footnotesize {\texttt{PSNR} ($\uparrow$)}}~ & - & - & - & 18.65 & 19.66 & 19.31 & \textbf{21.77}  \\  
         & & & ~{\footnotesize {\texttt{SSIM} ($\uparrow$)}}~ & - & - & - & 0.911 & 0.910 & 0.904 & \textbf{0.914} \\  [-0.5ex] 
        
\bottomrule  \\ [-3.6ex]  
    \end{tabular} 
}
\end{table}

%% file: sec/exp/fig_synth.tex
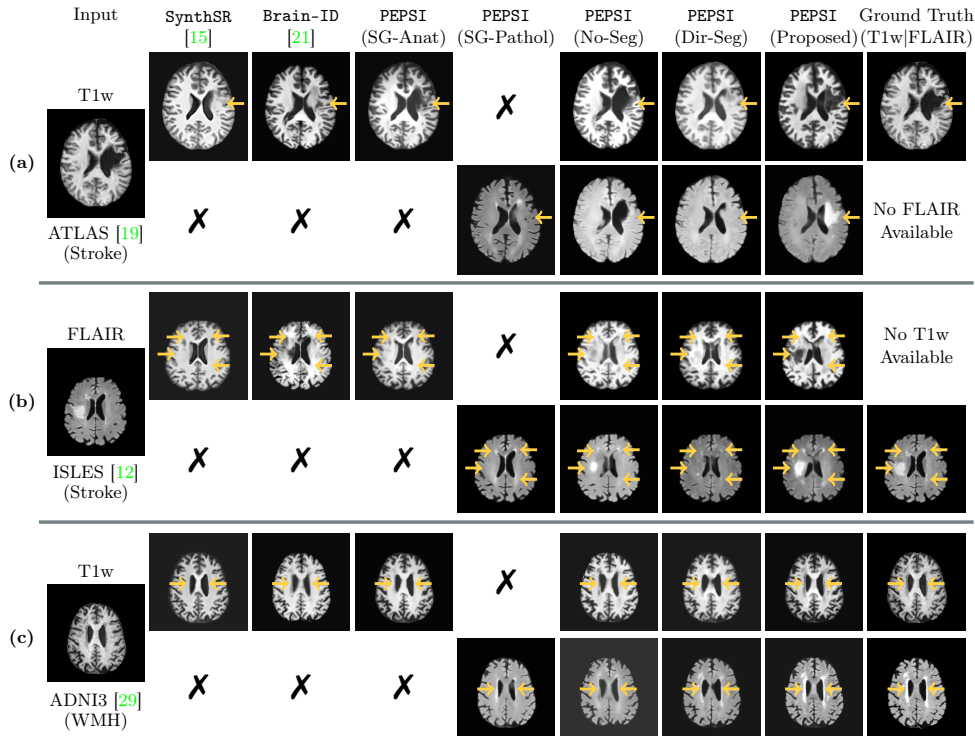
\begin{figure}[t]

\centering 

\resizebox{1.08\linewidth}{!}{
	\begin{tikzpicture}
        \hspace*{-0.9cm}
        
		\tikzstyle{myarrows}=[line width=0.8mm,draw=blue!50,-triangle 45,postaction={draw, line width=0.05mm, shorten >=0.02mm, -}]
		\tikzstyle{mylines}=[line width=0.8mm]
  

 
	\pgfmathsetmacro{\dx}{1.8}
	\pgfmathsetmacro{\shift}{-3.2}
	\node at (7.2+\dx+\shift+0.1*3, 6.6+0.15*3, 0.75*3) {\small Input};
	\node at (7.2+2*\dx+\shift+0.1*3, 6.55+0.15*3, 0.75*3) {\small \texttt{SynthSR}};
	\node at (7.2+2*\dx+\shift+0.1*3, 6.2+0.15*3, 0.75*3) {\small \cite{Iglesias2023SynthSRAP}};
	\node at (7.2+3*\dx+\shift+0.1*3, 6.6+0.15*3, 0.75*3) {\small \texttt{Brain-ID}}; 
	\node at (7.2+3*\dx+\shift+0.1*3, 6.2+0.15*3, 0.75*3) {\small \cite{Liu2023BrainID}}; 
	\node at (7.2+4*\dx+\shift+0.1*3, 6.6+0.15*3, 0.75*3) {\small \texttt{PEPSI}};
	\node at (7.2+4*\dx+\shift+0.1*3, 6.2+0.15*3, 0.75*3) {\small (SG-Anat)};
	\node at (7.2+5*\dx+\shift+0.1*3, 6.6+0.15*3, 0.75*3) {\small \texttt{PEPSI}};
	\node at (7.2+5*\dx+\shift+0.1*3, 6.2+0.15*3, 0.75*3) {\small (SG-Pathol)};
	\node at (7.2+6*\dx+\shift+0.1*3, 6.6+0.15*3, 0.75*3) {\small \texttt{PEPSI}};
	\node at (7.2+6*\dx+\shift+0.1*3, 6.2+0.15*3, 0.75*3) {\small (No-Seg)};
	\node at (7.2+7*\dx+\shift+0.1*3, 6.6+0.15*3, 0.75*3) {\small \texttt{PEPSI}};
	\node at (7.2+7*\dx+\shift+0.1*3, 6.2+0.15*3, 0.75*3) {\small (Dir-Seg)};
	\node at (7.2+8*\dx+\shift+0.1*3, 6.6+0.15*3, 0.75*3) {\small\color{black}\texttt{PEPSI}};
	\node at (7.2+8*\dx+\shift+0.1*3, 6.2+0.15*3, 0.75*3) {\small\color{black}({Proposed})};
	\node at (7.2+9*\dx+\shift+0.1*3, 6.6+0.15*3, 0.75*3) {\small Ground Truth};
	\node at (7.2+9*\dx+\shift+0.1*3, 6.2+0.15*3, 0.75*3) {\small (T1w$\vert$FLAIR)};

	\node at (7.2+\dx+\shift+0.1*3, 4.2-1.45+0.15*3, 0.75*3) {\small ATLAS~\cite{Liew2017ATLAS}};
	\node at (7.2+\dx+\shift+0.1*3, 3.8-1.4+0.15*3, 0.75*3) {\small (Stroke)};
	\node at (7.2+\dx+\shift+0.1*3, -0.-1.45+0.15*3, 0.75*3) {\small ISLES~\cite{Hernandez2022ISLES}};
	\node at (7.2+\dx+\shift+0.1*3, -0.4-1.4+0.15*3, 0.75*3) {\small (Stroke)};
	\node at (7.2+\dx+\shift+0.1*3, -4.2-1.45+0.2+0.15*3, 0.75*3) {\small ADNI3~\cite{Weiner2017TheAD}};
	\node at (7.2+\dx+\shift+0.1*3, -4.6-1.4+0.2+0.15*3, 0.75*3) {\small (WMH)};

	\node at (7.2+\dx+\shift+0.1*3, 4.2+1.+0.15*3, 0.75*3) {\small T1w};
	\node at (7.2+\dx+\shift+0.1*3, -0.+1.+0.15*3, 0.75*3) {\small FLAIR};
	\node at (7.2+\dx+\shift+0.1*3, -4.2+0.05+1.+0.15*3, 0.75*3) {\small T1w};

	\draw[line width = 0.06cm, color = flatgrey!150] (7.2-1+\dx+\shift+0.1*3, 1.9+0.15*3, 0.75*3) -- (7.2+1+9*\dx+\shift+0.1*3, 1.9+0.15*3, 0.75*3); 
	\draw[line width = 0.06cm, color = flatgrey!150] (7.2-1+\dx+\shift+0.1*3, 1.9-4.2+0.15*3, 0.75*3) -- (7.2+1+9*\dx+\shift+0.1*3, 1.9-4.2+0.15*3, 0.75*3); 

	\pgfmathsetmacro{\dy}{5}
  
	\node at (5.9+\dx+\shift+0.1*3, \dy-1+0.15*3, 0.75*3) {\small{\bf(a)}};
 
	\node at (7.2+\dx+\shift+0.1*3, \dy-1+0.15*3, 0.75*3) {\includegraphics[width=0.14\textwidth]{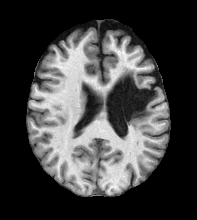}};
	\node at (7.2+2*\dx+\shift+0.1*3, \dy+0.15*3, 0.75*3) {\includegraphics[width=0.14\textwidth]{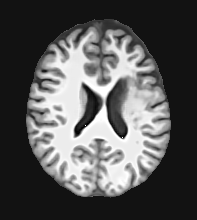}};
	\node at (7.2+3*\dx+\shift+0.1*3, \dy+0.15*3, 0.75*3) {\includegraphics[width=0.14\textwidth]{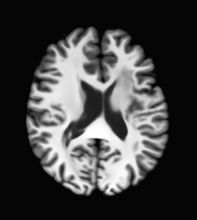}};
	\node at (7.2+4*\dx+\shift+0.1*3, \dy+0.15*3, 0.75*3) {\includegraphics[width=0.14\textwidth]{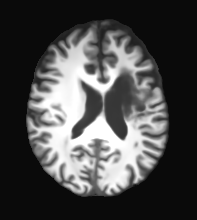}};  
	\node at (7.2+5*\dx+\shift+0.1*3, \dy+0.15*3, 0.75*3) {\LARGE\xmark};
	\node at (7.2+6*\dx+\shift+0.1*3, \dy+0.15*3, 0.75*3) {\includegraphics[width=0.14\textwidth]{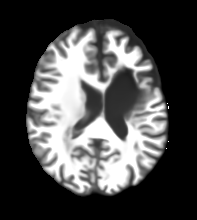}}; 
	\node at (7.2+7*\dx+\shift+0.1*3, \dy+0.15*3, 0.75*3) {\includegraphics[width=0.14\textwidth]{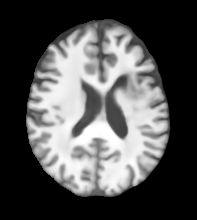}};  
	\node at (7.2+8*\dx+\shift+0.1*3, \dy+0.15*3, 0.75*3) {\includegraphics[width=0.14\textwidth]{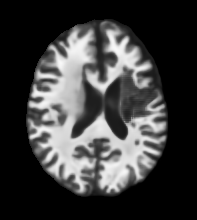}}; 
	\node at (7.2+9*\dx+\shift+0.1*3, \dy+0.15*3, 0.75*3) {\includegraphics[width=0.14\textwidth]{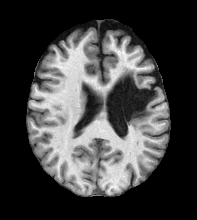}};


	\pgfmathsetmacro{\dy}{3}
 
	\node at (7.2+2*\dx+\shift+0.1*3, \dy+0.15*3, 0.75*3) {\LARGE\xmark};
	\node at (7.2+3*\dx+\shift+0.1*3, \dy+0.15*3, 0.75*3) {\LARGE\xmark};
	\node at (7.2+4*\dx+\shift+0.1*3, \dy+0.15*3, 0.75*3) {\LARGE\xmark}; 
	\node at (7.2+5*\dx+\shift+0.1*3, \dy+0.15*3, 0.75*3) {\includegraphics[width=0.14\textwidth]{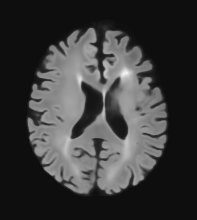}}; 
	\node at (7.2+6*\dx+\shift+0.1*3, \dy+0.15*3, 0.75*3) {\includegraphics[width=0.14\textwidth]{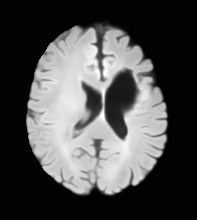}}; 
	\node at (7.2+7*\dx+\shift+0.1*3, \dy+0.15*3, 0.75*3) {\includegraphics[width=0.14\textwidth]{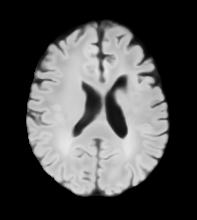}}; 
	\node at (7.2+8*\dx+\shift+0.1*3, \dy+0.15*3, 0.75*3) {\includegraphics[width=0.14\textwidth]{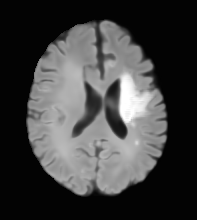}}; 
	\node at (7.2+9*\dx+\shift+0.1*3, \dy+0.15*3+0.2, 0.75*3) {\footnotesize No FLAIR}; 
	\node at (7.2+9*\dx+\shift+0.1*3, \dy+0.15*3-0.2, 0.75*3) {\footnotesize Available};

	\pgfmathsetmacro{\shiftx}{0.}
	\pgfmathsetmacro{\shifty}{8.4}
 
	\foreach \dx/\dy in {0/-8.25, 1.8/-8.25, 3.6/-8.25, 5.4/-10.25, 7.2/-8.25, 7.2/-10.25, 9./-8.25, 9./-10.25}
	{ 
        \draw[-To, draw=myyellow!150, line width=0.5mm] (8.4+\dx+\shiftx+0.1*3, 4.9+\dy+\shifty+0.15*3, 0.75*3) -- (8.1+\dx+\shiftx+0.1*3, 4.9+\dy+\shifty+0.15*3, 0.75*3);  
        } 

	\foreach \dx/\dy in {10.8/-8.25, 10.8/-10.25, 12.6/-8.25}
	{ 
        \draw[-To, draw=myyellow!150, line width=0.5mm] (8.4+\dx+\shiftx+0.1*3, 4.9+\dy+\shifty+0.15*3, 0.75*3) -- (8.1+\dx+\shiftx+0.1*3, 4.9+\dy+\shifty+0.15*3, 0.75*3);  
        }


	\pgfmathsetmacro{\dy}{0.8}
  
	\node at (5.9+\dx+\shift+0.1*3, \dy-1+0.15*3, 0.75*3) {\small{\bf(b)}};
 
	\node at (7.2+\dx+\shift+0.1*3, \dy-1+0.15*3, 0.75*3) {\includegraphics[width=0.14\textwidth]{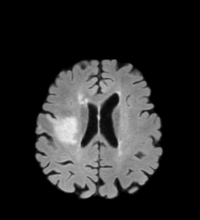}};
	\node at (7.2+2*\dx+\shift+0.1*3, \dy+0.15*3, 0.75*3) {\includegraphics[width=0.14\textwidth]{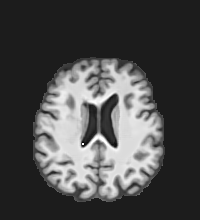}};
	\node at (7.2+3*\dx+\shift+0.1*3, \dy+0.15*3, 0.75*3) {\includegraphics[width=0.14\textwidth]{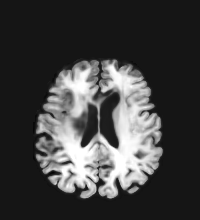}};
	\node at (7.2+4*\dx+\shift+0.1*3, \dy+0.15*3, 0.75*3) {\includegraphics[width=0.14\textwidth]{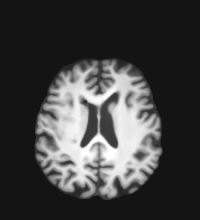}};  
	\node at (7.2+5*\dx+\shift+0.1*3, \dy+0.15*3, 0.75*3) {\LARGE\xmark}; 
	\node at (7.2+6*\dx+\shift+0.1*3, \dy+0.15*3, 0.75*3) {\includegraphics[width=0.14\textwidth]{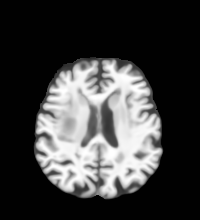}}; 
	\node at (7.2+7*\dx+\shift+0.1*3, \dy+0.15*3, 0.75*3) {\includegraphics[width=0.14\textwidth]{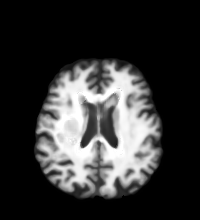}}; 
	\node at (7.2+8*\dx+\shift+0.1*3, \dy+0.15*3, 0.75*3) {\includegraphics[width=0.14\textwidth]{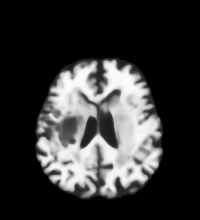}}; 
	\node at (7.2+9*\dx+\shift+0.1*3, \dy+0.15*3+0.2, 0.75*3) {\footnotesize No T1w};; 
	\node at (7.2+9*\dx+\shift+0.1*3, \dy+0.15*3-0.2, 0.75*3) {\footnotesize Available};;


	\pgfmathsetmacro{\dy}{-1.2}
   
	\node at (7.2+2*\dx+\shift+0.1*3, \dy+0.15*3, 0.75*3) {\LARGE\xmark};
	\node at (7.2+3*\dx+\shift+0.1*3, \dy+0.15*3, 0.75*3) {\LARGE\xmark};
	\node at (7.2+4*\dx+\shift+0.1*3, \dy+0.15*3, 0.75*3) {\LARGE\xmark};  
	\node at (7.2+5*\dx+\shift+0.1*3, \dy+0.15*3, 0.75*3) {\includegraphics[width=0.14\textwidth]{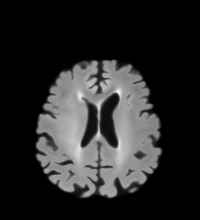}}; 
	\node at (7.2+6*\dx+\shift+0.1*3, \dy+0.15*3, 0.75*3) {\includegraphics[width=0.14\textwidth]{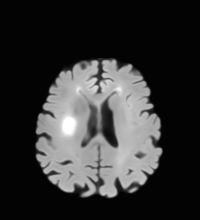}}; 
	\node at (7.2+7*\dx+\shift+0.1*3, \dy+0.15*3, 0.75*3) {\includegraphics[width=0.14\textwidth]{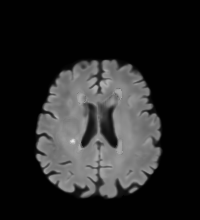}}; 
	\node at (7.2+8*\dx+\shift+0.1*3, \dy+0.15*3, 0.75*3) {\includegraphics[width=0.14\textwidth]{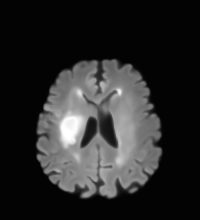}}; 
	\node at (7.2+9*\dx+\shift+0.1*3, \dy+0.15*3, 0.75*3) {\includegraphics[width=0.14\textwidth]{{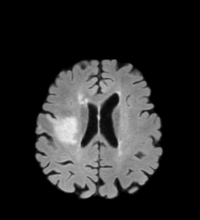}}};

	\pgfmathsetmacro{\shiftx}{-0.26}
	\pgfmathsetmacro{\shifty}{3.8}
 
	\foreach \dx/\dy in {0/-8.25, 1.8/-8.25, 3.6/-8.25, 5.4/-10.25, 7.2/-8.25, 7.2/-10.25, 9./-8.25, 9./-10.25}
	{ 
        \draw[-To, draw=myyellow!150, line width=0.5mm] (8.4+\dx+\shiftx+0.1*3, 5.42+\dy+\shifty+0.15*3, 0.75*3) -- (8.1+\dx+\shiftx+0.1*3, 5.42+\dy+\shifty+0.15*3, 0.75*3);
        \draw[-To, draw=myyellow!150, line width=0.5mm] (8.4+\dx+\shiftx+0.1*3, 4.9+\dy+\shifty+0.15*3, 0.75*3) -- (8.1+\dx+\shiftx+0.1*3, 4.9+\dy+\shifty+0.15*3, 0.75*3);
        \draw[-To, draw=myyellow!150, line width=0.5mm] (7.14+\dx+\shiftx+0.1*3, 5.1+\dy+\shifty+0.15*3, 0.75*3) -- (7.44+\dx+\shiftx+0.1*3, 5.1+\dy+\shifty+0.15*3, 0.75*3);  
        \draw[-To, draw=myyellow!150, line width=0.5mm] (7.3+\dx+\shiftx+0.1*3, 5.4+\dy+\shifty+0.15*3, 0.75*3) -- (7.6+\dx+\shiftx+0.1*3, 5.4+\dy+\shifty+0.15*3, 0.75*3);  
        } 

	\foreach \dx/\dy in {10.8/-8.25, 10.8/-10.25, 12.6/-10.25}
	{
        \draw[-To, draw=myyellow!150, line width=0.5mm] (8.4+\dx+\shiftx+0.1*3, 5.42+\dy+\shifty+0.15*3, 0.75*3) -- (8.1+\dx+\shiftx+0.1*3, 5.42+\dy+\shifty+0.15*3, 0.75*3);
        \draw[-To, draw=myyellow!150, line width=0.5mm] (8.4+\dx+\shiftx+0.1*3, 4.9+\dy+\shifty+0.15*3, 0.75*3) -- (8.1+\dx+\shiftx+0.1*3, 4.9+\dy+\shifty+0.15*3, 0.75*3);
        \draw[-To, draw=myyellow!150, line width=0.5mm] (7.14+\dx+\shiftx+0.1*3, 5.1+\dy+\shifty+0.15*3, 0.75*3) -- (7.44+\dx+\shiftx+0.1*3, 5.1+\dy+\shifty+0.15*3, 0.75*3);  
        \draw[-To, draw=myyellow!150, line width=0.5mm] (7.3+\dx+\shiftx+0.1*3, 5.4+\dy+\shifty+0.15*3, 0.75*3) -- (7.6+\dx+\shiftx+0.1*3, 5.4+\dy+\shifty+0.15*3, 0.75*3); 
        }


	\pgfmathsetmacro{\dy}{-3.35}
  
	\node at (5.9+\dx+\shift+0.1*3, \dy-1+0.15*3, 0.75*3) {\small{\bf(c)}};

	\node at (7.2+\dx+\shift+0.1*3, \dy+0.1-1+0.15*3, 0.75*3) {\includegraphics[width=0.14\textwidth]{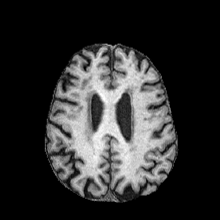}};
	\node at (7.2+2*\dx+\shift+0.1*3, \dy+0.15*3, 0.75*3) {\includegraphics[width=0.14\textwidth]{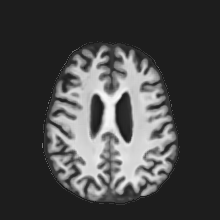}};
	\node at (7.2+3*\dx+\shift+0.1*3, \dy+0.15*3, 0.75*3) {\includegraphics[width=0.14\textwidth]{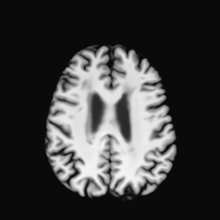}};
	\node at (7.2+4*\dx+\shift+0.1*3, \dy+0.15*3, 0.75*3) {\includegraphics[width=0.14\textwidth]{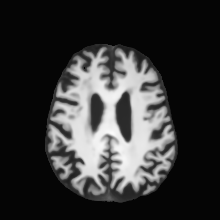}};  
	\node at (7.2+5*\dx+\shift+0.1*3, \dy+0.15*3, 0.75*3) {\LARGE\xmark};
	\node at (7.2+6*\dx+\shift+0.1*3, \dy+0.15*3, 0.75*3) {\includegraphics[width=0.14\textwidth]{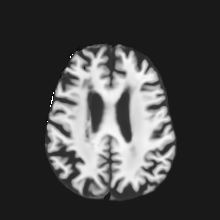}}; 
	\node at (7.2+7*\dx+\shift+0.1*3, \dy+0.15*3, 0.75*3) {\includegraphics[width=0.14\textwidth]{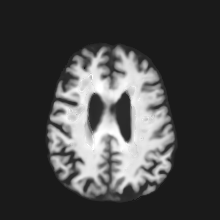}}; 
	\node at (7.2+8*\dx+\shift+0.1*3, \dy+0.15*3, 0.75*3) {\includegraphics[width=0.14\textwidth]{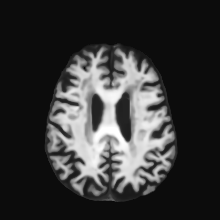}}; 
	\node at (7.2+9*\dx+\shift+0.1*3, \dy+0.15*3, 0.75*3) {\includegraphics[width=0.14\textwidth]{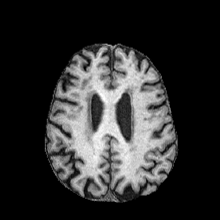}};


	\pgfmathsetmacro{\dy}{-5.2}
  
	\node at (7.2+2*\dx+\shift+0.1*3, \dy+0.15*3, 0.75*3) {\LARGE\xmark};
	\node at (7.2+3*\dx+\shift+0.1*3, \dy+0.15*3, 0.75*3) {\LARGE\xmark};
	\node at (7.2+4*\dx+\shift+0.1*3, \dy+0.15*3, 0.75*3) {\LARGE\xmark};  
	\node at (7.2+5*\dx+\shift+0.1*3, \dy+0.15*3, 0.75*3) {\includegraphics[width=0.14\textwidth]{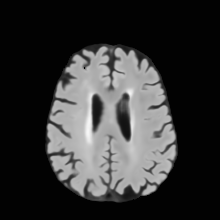}}; 
	\node at (7.2+6*\dx+\shift+0.1*3, \dy+0.15*3, 0.75*3) {\includegraphics[width=0.14\textwidth]{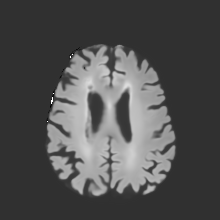}}; 
	\node at (7.2+7*\dx+\shift+0.1*3, \dy+0.15*3, 0.75*3) {\includegraphics[width=0.14\textwidth]{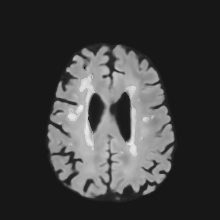}}; 
	\node at (7.2+8*\dx+\shift+0.1*3, \dy+0.15*3, 0.75*3) {\includegraphics[width=0.14\textwidth]{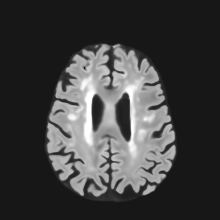}}; 
	\node at (7.2+9*\dx+\shift+0.1*3, \dy+0.15*3, 0.75*3) {\includegraphics[width=0.14\textwidth]{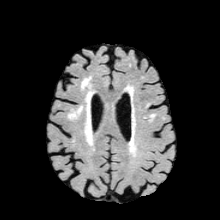}};

	\pgfmathsetmacro{\shiftx}{-0.05}
	\pgfmathsetmacro{\shifty}{-0.08}
 
	\foreach \dx/\dy in {0/-8.4, 1.8/-8.4, 3.6/-8.4, 5.4/-10.25, 7.2/-8.4, 7.2/-10.25, 9./-8.4, 9./-10.25}
	{
        \draw[-To, draw=myyellow!150, line width=0.5mm] (8.14+\dx+\shiftx+0.1*3, 5.1+\dy+\shifty+0.15*3, 0.75*3) -- (7.84+\dx+\shiftx+0.1*3, 5.1+\dy+\shifty+0.15*3, 0.75*3);
        \draw[-To, draw=myyellow!150, line width=0.5mm] (7.14+\dx+\shiftx+0.1*3, 5.1+\dy+\shifty+0.15*3, 0.75*3) -- (7.44+\dx+\shiftx+0.1*3, 5.1+\dy+\shifty+0.15*3, 0.75*3);  
        } 

	\foreach \dx/\dy in {10.8/-8.4, 10.8/-10.25, 12.6/-8.4, 12.6/-10.25}
	{
        \draw[-To, draw=myyellow!150, line width=0.5mm] (8.14+\dx+\shiftx+0.1*3, 5.1+\dy+\shifty+0.15*3, 0.75*3) -- (7.84+\dx+\shiftx+0.1*3, 5.1+\dy+\shifty+0.15*3, 0.75*3);
        \draw[-To, draw=myyellow!150, line width=0.5mm] (7.14+\dx+\shiftx+0.1*3, 5.1+\dy+\shifty+0.15*3, 0.75*3) -- (7.44+\dx+\shiftx+0.1*3, 5.1+\dy+\shifty+0.15*3, 0.75*3);  
        } 

	\end{tikzpicture}
	}  
	\vspace{-0.6cm}
\caption{Qualitative comparisons on anatomy and pathology ({\color{myyellow!180}\bf$\leftrightarrow$}) image synthesis.}
	 \label{fig: comp} 
\end{figure}

%% file: sec/exp/setup.tex

\vspace{-0.5cm}
\subsubsection{Datasets:}
\label{sec: datasets}
To cover a broader range of brain regions and pathologies, we train \texttt{PEPSI} on 1025 subjects from (\# of train/test cases): \textit{(i)} ADNI3~\cite{Weiner2017TheAD} (298/33), with $1~mm$ isotropic T1w and FLAIR pairs with WMH; \textit{(ii)}  ATLAS~\cite{Liew2017ATLAS} (590/65), with \textit{only} T1w and manually segmented stroke lesion for subacute/chronic stroke patients; \textit{(iii)} ISLES~\cite{Hernandez2022ISLES} (137/15), with \textit{only} FLAIR and stroke lesion segmentation for acute/subacute stroke patients. For pathology segmentation, we also test on   ISBI2015~\cite{Carass2017ISBI2015} and MSSEG2016~\cite{Commowick2021MSSEG}, comprising 21 and 15 WMH patients.

\vspace{-0.5cm}
\subsubsection{Metrics:}
\label{sec: metrics}
For image synthesis, we use \texttt{L1} distance, \texttt{PSNR}, and \texttt{SSIM} (structural similarity)~\cite{Liu2021SelfappearanceaidedDE}. For pathology segmentation, we use \texttt{Dice} scores~\cite{Billot2021SynthSegSO}.


\vspace{-0.5cm}
\subsubsection{Models:}
\label{sec: models}
We compare \texttt{PEPSI} with the state-of-the-art contrast-agnostic synthesis methods, \texttt{SynthSR}~\cite{Iglesias2023SynthSRAP} and \texttt{Brain-ID}~\cite{Liu2023BrainID}. We also evaluate \texttt{PEPSI}'s variants: \textit{(i-ii)} SG-Anat/Pathol: single-guidance from MR-RAGE/FLAIR; \textit{(iii-iv)} No/Dir-Seg: No/direct supervision from gold-standard pathology segmentations. 


\vspace{-0.5cm}
\subsubsection{Implementation Details:}
\label{sec: implement}
As a general feature representation model, \texttt{PEPSI} can use any backbone to extract features. For fairer comparison, we adopt the same five-level 3D UNet~\cite{Ronneberger2015UNetCN} as utilized in state-of-the-art models~\cite{Iglesias2023SynthSRAP,Liu2023BrainID} we compare with. Two linear layers are followed for anatomy and pathology image synthesis (\cref{sec: dual_supv}). The synthetic pathology-encoded data is of size $128^3$ (\cref{sec: generator}), with batch size as 4. We use AdamW optimizer, with a learning rate of $10^{-4}$ for the first 160,000 iterations and $10^{-5}$ until 240,000 iterations. We set $\lambda=1$ in \cref{eq: synth_loss}, and $\omega=0.1$ in \cref{eq: total_loss} for 100,000 iterations, and 1 afterward.

%% file: sec/exp/synthesis.tex
\subsection{Anatomy and Pathology Image Synthesis}
\vspace{-0.1cm}
\label{exp: pretrain}

As shown in \cref{tab: synth}, \texttt{PEPSI} achieves the best performance in synthesizing both T1w and FLAIR, across all datasets and pathologies. Notably, \texttt{PEPSI} exhibits superiority on single-modality datasets (ATLAS~\cite{Liew2017ATLAS}, ISLES~\cite{Hernandez2022ISLES}). Furthermore, \texttt{PEPSI} demonstrates strong \textit{robustness} against contrasts. For example, it maintains consistent scores for T1w synthesis on ADNI3~\cite{Weiner2017TheAD}, regardless of whether the input is T1w or FLAIR, whereas \texttt{SynthSR}~\cite{Iglesias2023SynthSRAP}, \texttt{Brain-ID}~\cite{Liu2023BrainID}, and other variants suffer from larger performance drops for FLAIR-to-T1w synthesis.

Thanks to the co-training and pathology-enhanced, contrast-agnostic learning, \texttt{PEPSI} can synthesize images that are \textit{not} present in the original datasets. \\
\noindent \cref{fig: comp}-{(a)}: \texttt{PEPSI} successfully synthesizes T1w and pathology-enhanced images based on T1w from ATLAS~\cite{Liew2017ATLAS}, for which ground truth FLAIR is not available. Remarkably, other models either \textit{cannot} estimate pathology-enhanced images, or struggle to \textit{accurately} capture and highlight (brighten) the areas of pathology.\\
\noindent \cref{fig: comp}-{(b)}: ISLES~\cite{Hernandez2022ISLES} only provides FLAIR and annotations for stroke lesions, yet \texttt{PEPSI}: \textit{(i)} accurately synthesizes T1w images with appropriately \textit{darkened} pathology regions inferred from the FLAIR input, and \textit{(ii)} is not constrained to the stroke lesions manually annotated by ISLES, but instead, captures (\textit{brightens}) all pathological regions including \textit{both} stroke lesions and WMH.

%% file: sec/exp/comp_curve.tex
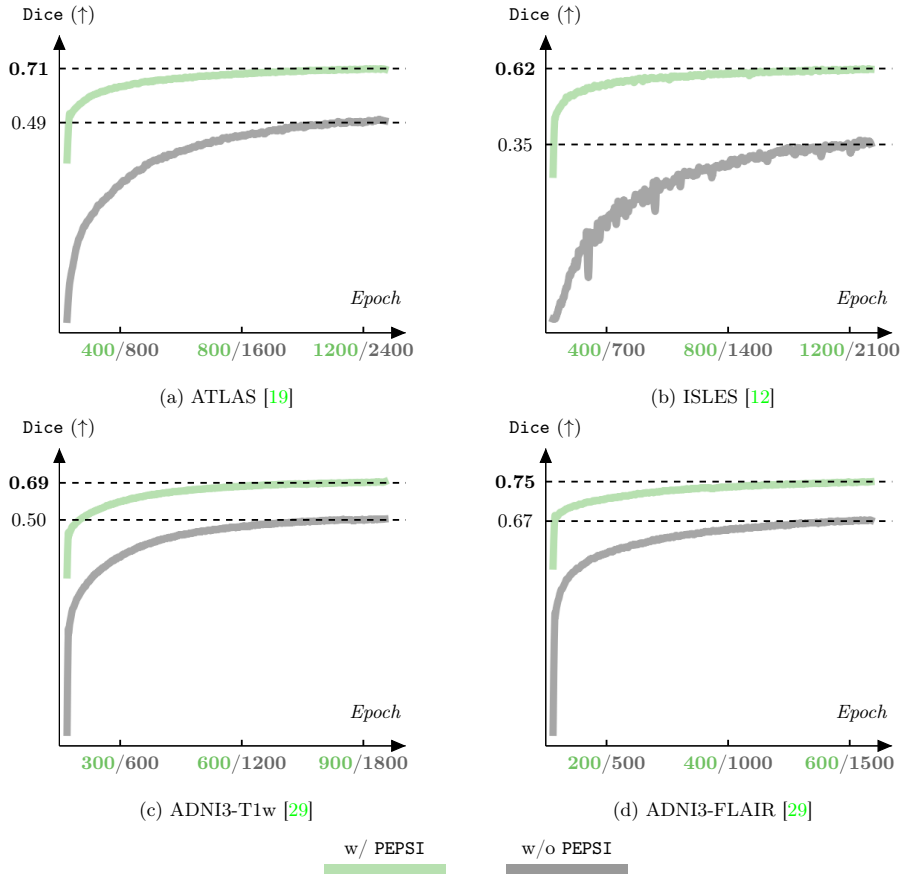
\begin{figure}[t]

\centering 

\resizebox{1\linewidth}{!}{
	\begin{tikzpicture}
        
		\tikzstyle{myarrows}=[line width=0.2mm,draw=blue!50,-triangle 45,postaction={draw, line width=0.05mm, shorten >=0.02mm, -}]
		\tikzstyle{mylines}=[line width=0.3mm]
  

 
	\pgfmathsetmacro{\shift}{-3.2}

	\node at (4.+\shift+0.1*3, 5.8+0.15*3, 0.75*3) {\includegraphics[width=0.50\textwidth]{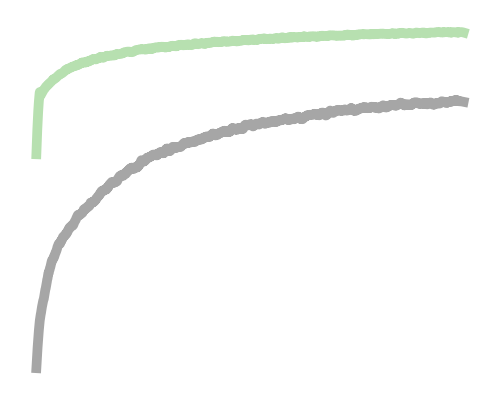}};
	\node at (4.+\shift+0.1*3, -1+0.15*3, 0.75*3) {\includegraphics[width=0.50\textwidth]{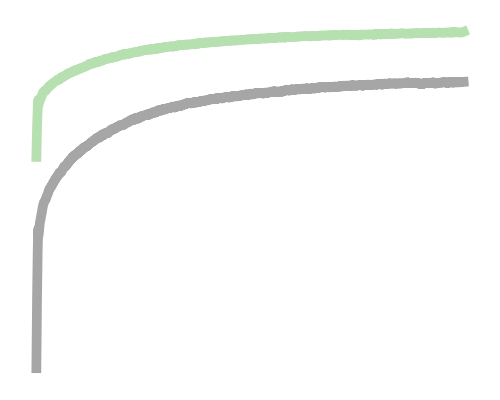}};
        
	\node at (12+\shift+0.1*3, 5.8+0.15*3, 0.75*3) {\includegraphics[width=0.50\textwidth]{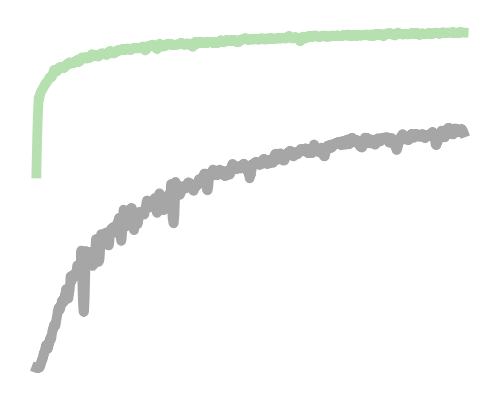}};
	\node at (12+\shift+0.1*3, -1+0.15*3, 0.75*3) {\includegraphics[width=0.50\textwidth]{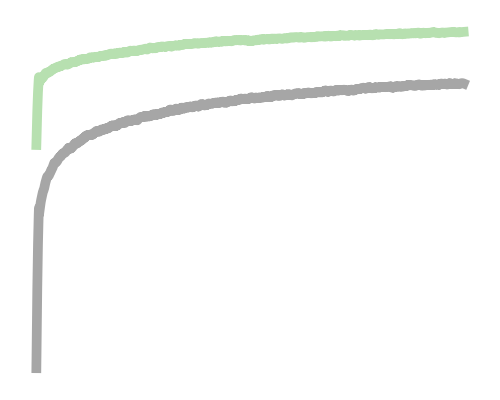}};


    \pgfmathsetmacro{\dx}{0}
    \pgfmathsetmacro{\dy}{0}
    
    \pgfmathsetmacro{\x}{-2.5}
    \pgfmathsetmacro{\y}{-3.7}
    
    \node[anchor=north] at (2.75+\x+\dx, -0.8+6.8+\y+\dy) {(a) ATLAS~\cite{Liew2017ATLAS}};
    \node[anchor=north] at (2.75+\x+\dx, -0.8+\y+\dy) {(c) ADNI3-T1w~\cite{Weiner2017TheAD}}; 
    \node[anchor=north] at (\x+\dx, 5.5+6.8+\y+\dy) {{\texttt{Dice} ($\uparrow$)}};
    \node[anchor=north] at (\x+\dx, 5.5+\y+\dy) {{\texttt{Dice} ($\uparrow$)}}; 
    
    \pgfmathsetmacro{\x}{-1.3}
    \node[anchor=north] at (2.75+6.8+\x+\dx, -0.8+6.8+\y+\dy) {(b) ISLES~\cite{Hernandez2022ISLES}}; 
    \node[anchor=north] at (2.75+6.8+\x+\dx, -0.8+\y+\dy) {(d) ADNI3-FLAIR~\cite{Weiner2017TheAD}};  
    
    \node[anchor=north] at (6.8+\x+\dx, 5.5+\y+\dy) {{\texttt{Dice} ($\uparrow$)}}; 
    \node[anchor=north] at (6.8+\x+\dx, 5.5+6.8+\y+\dy) {{\texttt{Dice} ($\uparrow$)}};

    \pgfmathsetmacro{\x}{-0.9}
    \pgfmathsetmacro{\dy}{-8.1} 
     
    \fill[fill=matcha] (2.75+\x+\dx, -0.8+6.8+\y+\dy) rectangle ++(2.0, 0.16);
    \node at (3.75+\x+\dx, -0.4+6.8+\y+\dy) {w/ \texttt{PEPSI}};

    \pgfmathsetmacro{\dx}{3}
    \fill[fill=gray!80] (2.75+\x+\dx, -0.8+6.8+\y+\dy) rectangle ++(2.0, 0.16);
    \node at (3.75+\x+\dx, -0.4+6.8+\y+\dy) {w/o \texttt{PEPSI}};


    \pgfmathsetmacro{\dx}{0}
    \pgfmathsetmacro{\dy}{0}
    
    \pgfmathsetmacro{\x}{-1.3}
    \pgfmathsetmacro{\y}{-3.7}

    \pgfmathsetmacro{\dx}{0}
    \pgfmathsetmacro{\a}{7.4}
    \pgfmathsetmacro{\b}{-3.1}
    \draw [mylines, dashed, color = black](6.8+\x+\dx, 3.75+\b+\dy) -- (5.7+6.8+\x+\dx, 3.75+\b+\dy);
    \node at (6.8-0.5+\x+\dx, 3.75\b+\dy) {\bf\small{0.75}};
    \draw [mylines, dashed, color = black](6.8+\x+\dx, 3.1+\b+\dy) -- (5.7+6.8+\x+\dx, 3.1+\b+\dy);
    \node at (6.8-0.5+\x+\dx, 3.1+\b+\dy) {\small{0.67}};

    \pgfmathsetmacro{\dx}{0}
    \pgfmathsetmacro{\dy}{0}
    
    \pgfmathsetmacro{\dx}{0}
    \pgfmathsetmacro{\a}{0.9}
    \pgfmathsetmacro{\b}{0.}
    \draw [mylines, dashed, color = black](6.8+\x+\dx, 6.5+0.95+\b+\dy) -- (5.7+6.8+\x+\dx, 6.5+0.95+\b+\dy);
    \node at (6.8-0.5+\x+\dx, 6.5+0.95+\b+\dy) {\bf\small{0.62}};
    \draw [mylines, dashed, color = black](6.8+\x+\dx, 6.2+\b+\dy) -- (5.7+6.8+\x+\dx, 6.2+\b+\dy);
    \node at (6.8-0.5+\x+\dx, 6.2+\b+\dy) {\small{0.35}}; 
    
    \pgfmathsetmacro{\dx}{-1.2}
    \pgfmathsetmacro{\a}{7.4}
    \pgfmathsetmacro{\b}{-3.1}
    \draw [mylines, dashed, color = black](\x+\dx, 3.75+6.8+\b+\dy) -- (5.7+\x+\dx, 3.75+6.8+\b+\dy);
    \node at (-0.5+\x+\dx, 3.75+6.8+\b+\dy) {\bf\small{0.71}};
    \draw [mylines, dashed, color = black](\x+\dx, 3.23+6.43+\b+\dy) -- (5.7+\x+\dx, 3.23+6.43+\b+\dy);
    \node at (-0.5+\x+\dx, 3.23+6.43+\b+\dy) {\small{0.49}};

    \pgfmathsetmacro{\dx}{-1.2}
    \pgfmathsetmacro{\a}{0.9}
    \pgfmathsetmacro{\b}{-3.1}
    \draw [mylines, dashed, color = black](\x+\dx, 3.73+\b+\dy) -- (5.7+\x+\dx, 3.73+\b+\dy);
    \node at (-0.5+\x+\dx, 3.73+\b+\dy) {\bf\small{0.69}};
    \draw [mylines, dashed, color = black](\x+\dx, 3.12+\b+\dy) -- (5.7+\x+\dx, 3.12+\b+\dy);
    \node at (-0.5+\x+\dx, 3.12+\b+\dy) {\small{0.50}};

    
	\foreach \dx/\dy in {-1.2/6.8}{ 
        \draw [mylines, color = black](1+\x+\dx, \y+\dy) -- (1+\x+\dx, 0.1+\y+\dy); 
        \node[] at (1+\x+\dx, -0.3\y+\dy) {\small{{\bf\color{matcha!200}400}/{\bf\color{gray!120}800}}};   
        \draw [mylines, color = black](3+\x+\dx, \y+\dy) -- (3+\x+\dx, 0.1+\y+\dy);  
        \node[] at (3+\x+\dx, -0.3\y+\dy) {\small{{\bf\color{matcha!200}800}/{\bf\color{gray!120}1600}}};    
        \draw [mylines, color = black](5+\x+\dx, \y+\dy) -- (5+\x+\dx, 0.1+\y+\dy); 
        \node[] at (5+\x+\dx, -0.3\y+\dy) {\small{{\bf\color{matcha!200}1200}/{\bf\color{gray!120}2400}}};  
    }

	\foreach \dx/\dy in {6.8/6.8}{ 
        \draw [mylines, color = black](1+\x+\dx, \y+\dy) -- (1+\x+\dx, 0.1+\y+\dy); 
        \node[] at (1+\x+\dx, -0.3\y+\dy) {\small{{\bf\color{matcha!200}400}/{\bf\color{gray!120}700}}};   
        \draw [mylines, color = black](3+\x+\dx, \y+\dy) -- (3+\x+\dx, 0.1+\y+\dy);  
        \node[] at (3+\x+\dx, -0.3\y+\dy) {\small{{\bf\color{matcha!200}800}/{\bf\color{gray!120}1400}}};    
        \draw [mylines, color = black](5+\x+\dx, \y+\dy) -- (5+\x+\dx, 0.1+\y+\dy); 
        \node[] at (5+\x+\dx, -0.3\y+\dy) {\small{{\bf\color{matcha!200}1200}/{\bf\color{gray!120}2100}}};  
    }

	\foreach \dx/\dy in {-1.2/0}{ 
        \draw [mylines, color = black](1+\x+\dx, \y+\dy) -- (1+\x+\dx, 0.1+\y+\dy); 
        \node[] at (1+\x+\dx, -0.3\y+\dy) {\small{{\bf\color{matcha!200}300}/{\bf\color{gray!120}600}}};   
        \draw [mylines, color = black](3+\x+\dx, \y+\dy) -- (3+\x+\dx, 0.1+\y+\dy);  
        \node[] at (3+\x+\dx, -0.3\y+\dy) {\small{{\bf\color{matcha!200}600}/{\bf\color{gray!120}1200}}};    
        \draw [mylines, color = black](5+\x+\dx, \y+\dy) -- (5+\x+\dx, 0.1+\y+\dy); 
        \node[] at (5+\x+\dx, -0.3\y+\dy) {\small{{\bf\color{matcha!200}900}/{\bf\color{gray!120}1800}}};  
    }

	\foreach \dx/\dy in {6.8/0}{ 
        \draw [mylines, color = black](1+\x+\dx, \y+\dy) -- (1+\x+\dx, 0.1+\y+\dy); 
        \node[] at (1+\x+\dx, -0.3\y+\dy) {\small{{\bf\color{matcha!200}200}/{\bf\color{gray!120}500}}};   
        \draw [mylines, color = black](3+\x+\dx, \y+\dy) -- (3+\x+\dx, 0.1+\y+\dy);  
        \node[] at (3+\x+\dx, -0.3\y+\dy) {\small{{\bf\color{matcha!200}400}/{\bf\color{gray!120}1000}}};    
        \draw [mylines, color = black](5+\x+\dx, \y+\dy) -- (5+\x+\dx, 0.1+\y+\dy); 
        \node[] at (5+\x+\dx, -0.3\y+\dy) {\small{{\bf\color{matcha!200}600}/{\bf\color{gray!120}1500}}};  
    }

	\foreach \dx/\dy in {-1.2/0, 6.8/0, -1.2/6.8, 6.8/6.8}
	{
    \draw [myarrows, color = black](\x+\dx, \y+\dy) -- (5.7+\x+\dx, \y+\dy);  
    \draw [myarrows, color = black](\x+\dx, \y+\dy) -- (\x+\dx, 4.9+\y+\dy); 
    
    \node[anchor=north] at (5.2+\x+\dx, 0.8+\y+\dy) {{\textit{Epoch}}};

}

	\end{tikzpicture}
	}  
	\vspace*{-0.35cm}
\caption{Training progresses of w/ \texttt{PEPSI} and w/o \texttt{PEPSI} for pathology segmentation. The horizontal (vertical) axis indicates training epochs ({\color{matcha!200}``w/ \texttt{PEPSI}'' epochs}~/~{\color{gray!120}``w/o \texttt{PEPSI}'' epochs}). 
Results are obtained by evaluating models collected throughout epochs.} 
 
	 \label{fig: comp_curve}
  
\end{figure}

%% file: sec/exp/tab_seg.tex
\begin{table}[ht]
    \caption{Average \texttt{Dice} scores ($\uparrow$) for pathology segmentation, w/o or w/ \texttt{PEPSI} pre-trained features. (Numbers in the parentheses denote the convergence/testing epochs ($\downarrow$); We directly test on ISBI2015 and MSSEG2016 using models trained from ADNI3.)} 
	\vspace*{-0.1cm}
    \label{tab: seg}
\resizebox{\linewidth}{!}{
\centering 
    \begin{tabular}{ccccccccc} 
       \toprule \\[-3ex] 
      \multicolumn{1}{c}{\multirow{2}{*}{\footnotesize\textbf{Model}}} & \multicolumn{1}{c}{\multirow{1}{*}{\footnotesize\textbf{ATLAS (Stroke)}}} &  \multicolumn{1}{c}{\multirow{1}{*}{\footnotesize\textbf{ISLES (Stroke)}}} & \multicolumn{2}{c}{\multirow{1}{*}{\footnotesize\textbf{ADNI3 (WMH)}}} & \multicolumn{2}{c}{\multirow{1}{*}{\footnotesize\textbf{ISBI2015 (WMH)}}} & \multicolumn{2}{c}{\multirow{1}{*}{\footnotesize\textbf{MSSEG2016 (WMH)}}} \\ [-0.5ex]
        \cmidrule(lr){4-5}
        \cmidrule(lr){6-7}
        \cmidrule(lr){8-9} 
        
          & {\footnotesize{T1w}} & {\footnotesize{FLAIR}} & \footnotesize{T1w} & \footnotesize{FLAIR} & \footnotesize{T1w} & \footnotesize{FLAIR} & \footnotesize{T1w} & \footnotesize{FLAIR}  \\ 
     \midrule\\[-3.8ex]

       \multicolumn{1}{c}{\multirow{2.5}{*}{\footnotesize{w/o \texttt{PEPSI}}}} & \multicolumn{1}{c}{\multirow{2}{*}{\thead{\footnotesize{$0.49 \pm 0.14$}\\\footnotesize{(2500)}}}} & \multicolumn{1}{c}{\multirow{2}{*}{\thead{\footnotesize{$0.35 \pm 0.13$}\\\footnotesize{(2000)}}}}& \multicolumn{1}{c}{\multirow{2}{*}{\thead{\footnotesize{$0.50 \pm 0.15$}\\\footnotesize{(1600)}}}}& \multicolumn{1}{c}{\multirow{2}{*}{\thead{\footnotesize{$0.67 \pm 0.13$}\\\footnotesize{(1500)}}}}& \multicolumn{1}{c}{\multirow{2}{*}{\thead{\footnotesize{$0.21 \pm 0.05$}\\\footnotesize{(1600)}}}}& \multicolumn{1}{c}{\multirow{2}{*}{\thead{\footnotesize{$0.39 \pm 0.15$}\\\footnotesize{(1500)}}}}& \multicolumn{1}{c}{\multirow{2}{*}{\thead{\footnotesize{$0.24 \pm 0.09$}\\\footnotesize{(1600)}}}}& \multicolumn{1}{c}{\multirow{2}{*}{\thead{\footnotesize{$0.31 \pm 0.10$}\\\footnotesize{(1500)}}}} \\ 

        \\  [1ex]

       \hline  
     
       \multicolumn{1}{c}{\multirow{2.5}{*}{\footnotesize{w/ \texttt{PEPSI}}}} & \multicolumn{1}{c}{\multirow{2}{*}{\thead{\footnotesize{${\bf{0.71}}\pm 0.22$}\\\footnotesize{(\textbf{1000})}}}} & \multicolumn{1}{c}{\multirow{2}{*}{\thead{\footnotesize{${\bf{0.62}} \pm 0.27$}\\\footnotesize{(\textbf{500})}}}} & \multicolumn{1}{c}{\multirow{2}{*}{\thead{\footnotesize{${\bf{0.69}} \pm 0.12$}\\\footnotesize{(\textbf{800})}}}} & \multicolumn{1}{c}{\multirow{2}{*}{\thead{\footnotesize{${\bf{0.75}} \pm 0.10$}\\\footnotesize{(\textbf{500})}}}} & \multicolumn{1}{c}{\multirow{2}{*}{\thead{\footnotesize{${\bf{0.34}} \pm 0.06$}\\\footnotesize{(\textbf{800})}}}} & \multicolumn{1}{c}{\multirow{2}{*}{\thead{\footnotesize{${\bf{0.57}} \pm 0.15$}\\\footnotesize{(\textbf{500})}}}} & \multicolumn{1}{c}{\multirow{2}{*}{\thead{\footnotesize{${\bf{0.38}} \pm 0.10$}\\\footnotesize{(\textbf{800})}}}} & \multicolumn{1}{c}{\multirow{2}{*}{\thead{\footnotesize{${\bf{0.45}} \pm 0.11$}\\\footnotesize{(\textbf{500})}}}} \\ 
       
         \\ [1ex] 
\bottomrule  \\ [-3.6ex]  
    \end{tabular} 
}
\end{table}

%% file: sec/exp/fig_seg.tex
\begin{figure}[ht]
\centering 

\resizebox{0.9\linewidth}{!}{
	\begin{tikzpicture}
        
		\tikzstyle{myarrows}=[line width=0.8mm,draw=blue!50,-triangle 45,postaction={draw, line width=0.05mm, shorten >=0.02mm, -}]
		\tikzstyle{mylines}=[line width=0.8mm]
  


	\pgfmathsetmacro{\shift}{-3.2}
	\node at (5.2+\shift+0.1*3, 6.2+0.15*3, 0.75*3) {Input}; 
	\node at (7.7+\shift+0.1*3, 6.2+0.15*3, 0.75*3) {w/o \texttt{PEPSI}};
	\node at (10.2+\shift+0.1*3, 6.2+0.15*3, 0.75*3) {w/ \texttt{PEPSI}};
	\node at (12.7+\shift+0.1*3, 6.2+0.15*3, 0.75*3) {Gold-standard};


	\pgfmathsetmacro{\dy}{4.8}
 
	\node at (5.2+\shift+0.1*3, \dy+0.15*3, 0.75*3) {\includegraphics[width=0.16\textwidth]{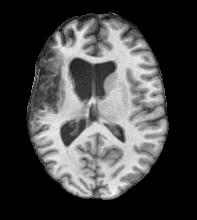}};
	\node at (7.7+\shift+0.1*3, \dy+0.15*3, 0.75*3) {\includegraphics[width=0.16\textwidth]{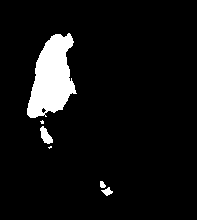}};
	\node at (10.2+\shift+0.1*3, \dy+0.15*3, 0.75*3) {\includegraphics[width=0.16\textwidth]{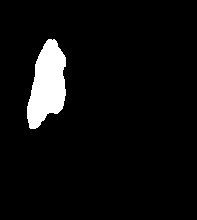}}; 
	\node at (12.7+\shift+0.1*3, \dy+0.15*3, 0.75*3) {\includegraphics[width=0.16\textwidth]{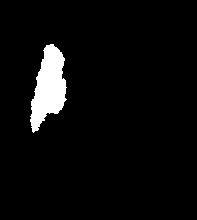}}; 
 
	\node at (8.95+\shift+0.1*3, \dy-1.3-0.1+0.15*3, 0.75*3) {(a) ATLAS~\cite{Liew2017ATLAS} (T1w)};

	\pgfmathsetmacro{\dy}{2.05}
 
	\node at (5.2+\shift+0.1*3, \dy+0.15*3, 0.75*3) {\includegraphics[width=0.16\textwidth]{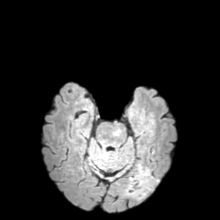}};
	\node at (7.7+\shift+0.1*3, \dy+0.15*3, 0.75*3) {\includegraphics[width=0.16\textwidth]{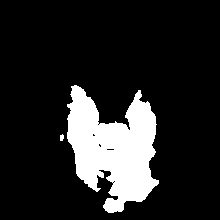}};
	\node at (10.2+\shift+0.1*3, \dy+0.15*3, 0.75*3) {\includegraphics[width=0.16\textwidth]{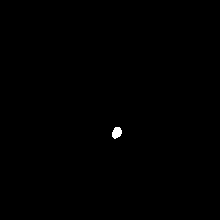}};
	\node at (12.7+\shift+0.1*3, \dy+0.15*3, 0.75*3) {\includegraphics[width=0.16\textwidth]{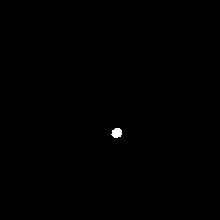}}; 

	\node at (8.95+\shift+0.1*3, \dy-1.3+0.15*3, 0.75*3) {(b) ISLES~\cite{Hernandez2022ISLES} (FLAIR)};

	\pgfmathsetmacro{\dy}{-0.6}
 
	\node at (5.2+\shift+0.1*3, \dy+0.15*3, 0.75*3) {\includegraphics[width=0.16\textwidth]{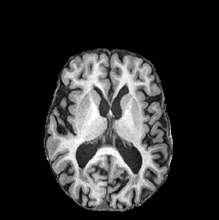}};
	\node at (7.7+\shift+0.1*3, \dy+0.15*3, 0.75*3) {\includegraphics[width=0.16\textwidth]{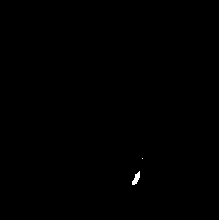}};
	\node at (10.2+\shift+0.1*3, \dy+0.15*3, 0.75*3) {\includegraphics[width=0.16\textwidth]{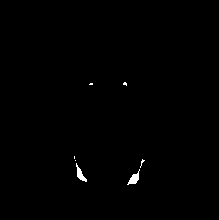}};
	\node at (12.7+\shift+0.1*3, \dy+0.15*3, 0.75*3) {\includegraphics[width=0.16\textwidth]{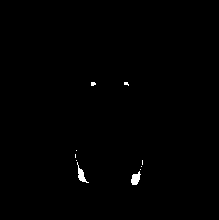}};

	\node at (8.95+\shift+0.1*3, \dy-1.3+0.15*3, 0.75*3) {(c) ADNI3~\cite{Weiner2017TheAD} (T1w)};

	\pgfmathsetmacro{\dy}{-3.4}
 
	\node at (5.2+\shift+0.1*3, \dy+0.15*3, 0.75*3) {\includegraphics[width=0.16\textwidth]{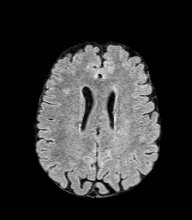}};
	\node at (7.7+\shift+0.1*3, \dy+0.15*3, 0.75*3) {\includegraphics[width=0.16\textwidth]{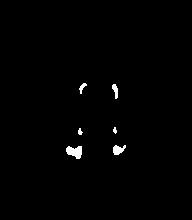}};
	\node at (10.2+\shift+0.1*3, \dy+0.15*3, 0.75*3) {\includegraphics[width=0.16\textwidth]{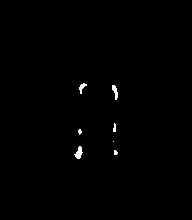}};
	\node at (12.7+\shift+0.1*3, \dy+0.15*3, 0.75*3) {\includegraphics[width=0.16\textwidth]{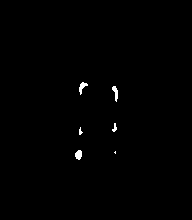}};

	\node at (8.95+\shift+0.1*3, \dy-1.3+0.1*3, 0.75*3) {(d) ADNI3~\cite{Weiner2017TheAD} (FLAIR)};

	\pgfmathsetmacro{\dy}{-6.35}
 
	\node at (5.2+\shift+0.1*3, \dy+0.15*3, 0.75*3) {\includegraphics[width=0.16\textwidth]{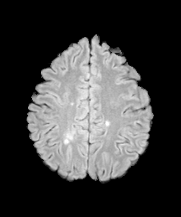}};
	\node at (7.7+\shift+0.1*3, \dy+0.15*3, 0.75*3) {\includegraphics[width=0.16\textwidth]{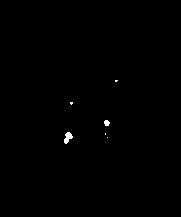}};
	\node at (10.2+\shift+0.1*3, \dy+0.15*3, 0.75*3) {\includegraphics[width=0.16\textwidth]{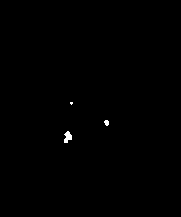}};
	\node at (12.7+\shift+0.1*3, \dy+0.15*3, 0.75*3) {\includegraphics[width=0.16\textwidth]{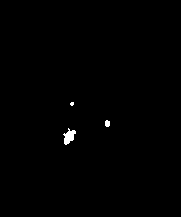}};

	\node at (8.95+\shift+0.1*3, \dy-1.47+0.15*3, 0.75*3) {(e) ISBI2015~\cite{Carass2017ISBI2015} (FLAIR)};

	\pgfmathsetmacro{\dy}{-9.36}
 
	\node at (5.2+\shift+0.1*3, \dy+0.15*3, 0.75*3) {\includegraphics[width=0.16\textwidth]{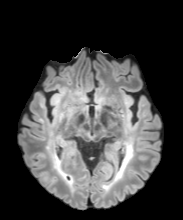}};
	\node at (7.7+\shift+0.1*3, \dy+0.15*3, 0.75*3) {\includegraphics[width=0.16\textwidth]{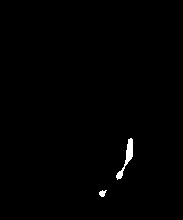}};
	\node at (10.2+\shift+0.1*3, \dy+0.15*3, 0.75*3) {\includegraphics[width=0.16\textwidth]{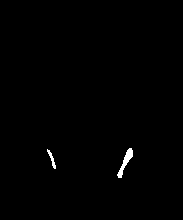}};
	\node at (12.7+\shift+0.1*3, \dy+0.15*3, 0.75*3) {\includegraphics[width=0.16\textwidth]{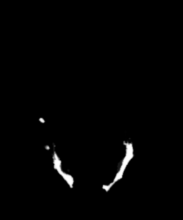}};

	\node at (8.95+\shift+0.1*3, \dy-1.47+0.15*3, 0.75*3) {(f) MSSEG2016~\cite{Commowick2021MSSEG} (FLAIR)};


	\end{tikzpicture}
	}  
	\vspace{-0.3cm}
\caption{Qualitative comparisons on downstream pathology segmentation, w/o or w/ \texttt{PEPSI} pre-trained features.} 

	 \label{fig: seg} 
\end{figure}

%% file: sec/exp/segment.tex
\subsection{Pathology Segmentation}
\vspace{-0.1cm}
\label{exp: downstream}

In \cref{exp: pretrain}, we validate \texttt{PEPSI}'s superiority in synthesizing pathology-enhanced images under various contrasts, providing voxel-level information that is \textit{not} confined to particular pathology types, but contains comprehensive information on anomalies. We further illustrate the \textit{efficiency} and \textit{effectiveness} of \texttt{PEPSI} features for downstream pathology segmentations that target a specific pathology. 

To this end, we compare the following two models trained on each dataset and contrast, \textit{(i)} starting from random weights (w/o \texttt{PEPSI}), and \textit{(ii)} fine-tuned from \texttt{PEPSI} pre-trained weights (w/ \texttt{PEPSI}). For ATLAS~\cite{Liew2017ATLAS}, ISLES~\cite{Hernandez2022ISLES}, and ADNI3~\cite{Weiner2017TheAD}, both models are trained and tested on their respective training and testing sets. Since ISBI2015~\cite{Carass2017ISBI2015} and MSSEG2016~\cite{Commowick2021MSSEG} datasets contain only 21 and 15 WMH cases, respectively, we directly evaluate the trained models from ADNI3 (WMH)~\cite{Weiner2017TheAD} on all available cases in these datasets. Note that although \texttt{PEPSI} has undergone pre-training on synthetic data using anatomy labels and pathology probability maps from the training sets of ATLAS~\cite{Liew2017ATLAS} and ISLES~\cite{Hernandez2022ISLES} (\cref{sec: datasets}), it has \textit{not} been exposed to any real image during the pre-training stage. 

As shown in \cref{fig: comp_curve}, utilizing \texttt{PEPSI}'s pre-trained features largely reduces the convergence time (by $\approx$ 60\% on average). More importantly, quantitative comparisons in \cref{tab: seg} demonstrate that \texttt{PEPSI} features yield higher \texttt{Dice} scores compared with models trained from scratch (i.e., w/o \texttt{PEPSI}) on all testing pathologies, contrasts and datasets. 
Furthermore, when directly tested on the two small datasets (ISBI2015 and MSSEG2016), \texttt{PEPSI} exhibits superior generalizability compared to models trained without \texttt{PEPSI} pre-trained features. Qualitative comparisons of pathology segmentations between w/o and w/ \texttt{PEPSI} on all five experimented datasets can be found in \cref{fig: seg}.

%% file: sec/con.tex
\section{Conclusion}
\label{sec: con} 
We introduced \texttt{PEPSI}, the first pathology-enhanced, contrast-agnostic feature representation learning approach for brain MRI. Trained on synthetic data featuring diverse contrasts, anomaly intensities and shapes, \texttt{PEPSI} exhibits remarkable robustness and accurately captures anomalies beyond the specific, manually annotated pathology, regardless of MR contrasts. We demonstrated \texttt{PEPSI}'s performance on anatomy and pathology image synthesis, covering T1w and FLAIR with stroke lesions and WMH, and further showcased the efficiency and effectiveness of \texttt{PEPSI} features for downstream pathology segmentation on five public datasets. We believe \texttt{PEPSI} will pave the way for the exciting future of contrast-agnostic pathology representations for heterogeneous, real-world brain MRI -- enabling studies of diverse brain diseases with large clinical MRI datasets.